\documentclass[showpacs,preprintnumbers]{revtex4}
\usepackage{graphicx}
\usepackage{subfig}
\usepackage{color}
\usepackage{amsmath}
\usepackage{amssymb}
\usepackage{mleftright}
\mleftright
\def\bc{\begin{center}}
\def\ec{\end{center}}

\def\beq{\begin{equation}}
\def\eeq{\end{equation}}

\begin{document}

\title{Time evolution of the quantum entanglement between $N$ qubits due to dynamical Lamb effect in the presence of dissipation}
\author{Mirko Amico$^{1,2}$, Oleg L. Berman$^{1,2}$ and Roman Ya. Kezerashvili$^{1,2}$}
\affiliation{\mbox{$^{1}$Physics Department, New York City College
of Technology, The City University of New York,} \\
Brooklyn, NY 11201, USA \\
\mbox{$^{2}$The Graduate School and University Center, The
City University of New York,} \\
New York, NY 10016, USA}
\date{\today}

\date{\today}

\begin{abstract}

A theoretical framework to investigate the time evolution of the quantum entanglement due to the dynamical Lamb effect between $N$ superconducting qubits coupled to a coplanar waveguide in the presence of different sources of dissipation is developed. 
We quantitatively analyze the case of $N=2$ and $3$ qubits under the assumptions of single switching of the coupling and absence of dissipation within a perturbative approach. The same systems are analyzed for the general case of periodic switching of the coupling in the presence of dissipation via numerical calculations.
Different measures of entanglement compatible with mixed states are adopted. It is demonstrated that the different measures show different level of details of the latter. 
The concurrence and the negativity are obtained in the two qubits case, the three-$\pi$ and the negativity in the three qubits case.
It is shown that time-dependent Greenberger-Horne-Zeilinger states can be created even in presence of dissipation.
To maximize the quantum entanglement between the qubits, the effects of tuning several parameters of the system are investigated.

\end{abstract}


\maketitle

\section{Introduction}
\label{intro}

Recent experiments in circuit quantum electrodynamics have demonstrated the possibility of probing quantum vacuum phenomena which have no classical analog. These phenomena arise when the vacuum is perturbed and virtual fluctuations are converted into real particles. One example is the dynamical Casimir effect (DCE), a dynamical version of the Casimir effect \cite{casimir}. The latter results in the creation of real photons from the vacuum electromagnetic field of a cavity through the nonadiabatic modulation of its boundary conditions. The DCE was predicted by Moore in Ref. \cite{moore} and it was recently observed in experiments with superconducting circuit devices \cite{wilson} and Josephson metamaterials \cite{lahte}. Another quantum vacuum phenomenon which also arises in the nonadiabatic regime is the dynamical Lamb effect (DLE), which was first described in Ref. \cite{DLE}. One can think of it in the following way, an atom in a cavity is characterized by a certain Lamb shift which depends on the vacuum electromagnetic field of the cavity. Depending on the size of the cavity, only a certain set of modes of the electromagnetic field are allowed. By changing the size of the cavity nonadiabatically, the set of allowed mode of the electromagnetic field in the cavity suddendly changes. Therefore, the atom experiences an instantaneous change of its Lamb shift, which in turn leads to its parametric excitation. In Ref. \cite{DLE} a first proposal on how to give rise to the DLE for atoms in a cavity was given. However, it is quite difficult to implement such hypothetical setup with real atoms and cavities. In Refs. \cite{shapiro, zhukov}, it was proposed to use superconducting qubits (as atoms) coupled to a coplanar waveguide (as cavity) to realize the nonadiabatic change of boubdary conditions. To be nonadiabatic, the change in boundary condition of the cavity has to happen in a time $\tau$ smaller than any parameter of the system with dimensions of time, $\tau \ll E_0^{-1}$ and $\tau \ll \omega_c^{-1} $, where $E_0$ is the transition energy of the qubits (here and through the rest of the article, we take $\hbar = 1$) and $\omega_c$ is the frequency of the cavity photons.



In Ref. \cite{berman} it was shown that for the case of two qubits coupled to a nonstationary cavity, the DCE and the DLE generate quantum entanglement between the qubits. The work was taken a step further in Ref. \cite{amico}, where the case of three qubits coupled to a nonstationary cavity was treated. For this case, it was found that the DCE/DLE can lead to the simultaneous entanglement of all three qubits, forming a Greenberger-Horne-Zeilinger (GHZ) state \cite{ghz,ghz2}. However, for the case of a nonstationary cavity, it is not possible to isolate the contribution of the DLE to the quantum entanglement from the contribution of the DCE. This difficulty can be overcome by following a different approach, presented in Ref. \cite{shapiro}. If the cavity is taken to be stationary while the qubit/cavity coupling is modulated nonadiabatically, the DLE becomes the only contribution to the excitation of qubits and cavity photons. The possibility of turning on/off the qubit/cavity coupling was experimentally demonstrated in Ref. \cite{lu} and its fast tunability shown in Ref. \cite{chen}. Both features can be achieved by modulating the magnetic flux passing through an auxillary superconducting quantum interference device (SQUID). The problem of studying the time evolution of the quantum entanglement due to the DLE was first presented in Ref. \cite{remizov}, where the dissipative evolution of a system of two qubits coupled to a cavity through nonadiabatic modulation of its coupling is considered. It was demonstrated that by switching the qubit/cavity coupling on and off periodically, the concurrence saturates at a fraction of its maximum value. Thus showing that the DLE can be used to generate quantum entanglement between qubits reliably.

We present a study of a system of $N$ qubits coupled to a common cavity, where the qubit/cavity coupling is suddenly and periodically switched on/off in the presence of dissipation that extends previous work on the entanglement created by the DLE. The tunable coupling, in contrast to the nonstationary cavity, allows to isolate the DLE as the only source of quantum entanglement between the qubits. Furthermore, the suppression of the DCE has a positive effect on the entanglement of the qubits, as photons created by the DCE tend to destroy quantum correlations between the qubits. Our interest in a system of $N$ qubits stems from the possible application of this model to Josephson metamaterials \cite{shulga, macha, zagoskin, lahte} and from the ongoing effort in making a quantum computer with more and more superconducting qubits. For instance, in Ref. \cite{song} ten superconducting transmon qubits were coupled to a common resonator. A realistic description of a system of $N$ qubits coupled to a cavity requires careful consideration of dissipative effects. Studying the interplay of dissipation and driving allows to determine the steady-state properties of the system. We propose a theoretical approach to describe the DLE in a Josephson metamaterial and the entanglement that is consecuently generated. In particular, we are interested in the dissipative evolution of the quantum entanglement between the qubits. In previous work \cite{berman, amico}, where dissipation was not considered, we used the concurrence \cite{wootters} and the tangle \cite{coffman} to quantify the two and three-way entanglement of the pure states of the qubits in the Hilbert space. Here, we adopt the density matrix formalism to describe the state of the system in the presence of dissipation, which can be an incoherent mixture of pure states or a mixed state in Liouville space. In this way, it's possible to study the time evolution of the system taking into account its interaction with the environment. In the following analysis we quantify the entanglement in a system of two qubits by using the concurrence \cite{wootters}, the mutual information \cite{plenio} and the negativity \cite{vidal} and for the three qubits case we use the negativity and the three-$\pi$ \cite{ou}. The consideration of different measures of entanglement highlights different details of the entanglement in the system, revealing features which are not captured by one single measure. Moreover, we investigate the dependence of these quantities on the system's parameter to find the values which maximize the quantum entanglement of the system. In particular, the dependence on the following parameters is studied: the frequency of switching of the qubit/cavity coupling, the frequency of the resonant mode of the cavity and the cavity dissipation rate.

From the numerical calculations of the two qubits case, we find that when the qubit/cavity coupling is turned on/off nonadiabatically and periodically at a frequency equal to the sum frequency of the transition frequencies of the qubits, it is possible to reach the maximum value of the measures of entanglement used periodically with time. Two different scenarios are possible depending on the value of the cavity and qubit dissipation rates. If the cavity dissipation rate is low, the measures of entanglement reach their maximum value periodically with time, if the latter is high, they saturate at a fraction of the maximum value.  
For the case of three qubits with the same transition frequency, we find that driving the qubit/cavity coupling at a frequency equal to the sum frequency of the transition frequencies of the qubits maximizes the simultaneous entanglement of three qubits. This is indicated by the three-$\pi$, which reaches its maximum value periodically with time when the qubit/cavity coupling is under modulation. Therefore, time-dependent GHZ states can be created even in presence of dissipation. The importance of maximally entangled states, or GHZ states, of three and more qubits comes from the fact that they can be used to test the validity of quantum mechanics (GHZ theorem \cite{ghz, ghz2}). Furthermore, quantum error correction codes rely on the ability to produce entangled states to protect quantum information from unwanted errors. A simple quantum error correction code for superconducting qubits, which requires to encode a qubit into an entangled states of three qubits, was presented in Ref. \cite{reed}. Alternatively, we find that when the three qubits have different transition frequencies, the driving of the coupling selectively entangles the pair of qubits for which the sum of the transition frequencies matches the frequency of the driving of the qubit/cavity coupling. The ability to perform entangling two qubits gates is fundamental to form an elementary set of quantum gates for quantum computation, which allows to carry out any unitary operation as proven in Ref. \cite{barenco}.



The article is organized in the following way. In Sec. \ref{ment}, different measures of entanglement are introduced as a way to quantify the entanglement between two, three and $N$ qubits. Sec. \ref{nq} describes the methods of analysis valid for the general $N$ qubits case. The Hamiltonian of the system is specified and an analytical and numerical approach to find the time evolution of the system is proposed. We consider the particular case of two and three qubits in Sec. \ref{2q} and \ref{3q}, respectively. 
We find the time evolution of the quantum entanglement between the qubits, in the framework of a perturbative approach for the case of no dissipation and time independent perturbation, and within a numerical approach for the case of periodic switching of the coupling in the presence of dissipation.
To find the optimal values of the parameters of the system which maximize the quantum entanglement, the values of various parameters are changed over an experimentally accessible range. The discussion of the results is presented in Sec. \ref{results} and the conclusions follow in Sec. \ref{conclusions}.

\section{Measures of quantum entanglement}
\label{ment}

We are interested in quantifying the simultaneous entanglement between all qubits in the system. In general, the problem of detecting and quantifying the multipartite entanglement in a system of $N$ qubits with mixed states, is a very challenging one. Refs. \cite{horodecki, plenio} contain a review of possible candidates to be a measure of the entanglement, however each of them emphasizes a particular aspect of entanglement and, as of today, no particular one has become the standard. In order to define an entanglement measure, the quantity has to satisfy the following minimal set of requirements, first stated in Ref. \cite{vedral}: i. it is a function of positive values; ii. it is zero for separable state; iii. its value does not increase under local operations and classical communication (LOCC). Further postulates such as additivity and continuity can be made to construct a measure with desirable properties but are not strictly necessary.

Below, we present the measures of quantum entanglement used for the case of two, three and $N$ qubits. The different measures of entanglement show different level of details of the latter. Therefore, the use of multiple measures helps us draw a richer picture of the features of the quantum entanglement between the qubits.

\subsection{Quantum entanglement of two qubits}
We adopt different measures of entanglement to quantify its time evolution. One of the measures of entanglement that we use for the two qubit case is the concurrence $C$, introduced in Ref. \cite{wootters}, which is valid both for pure and mixed state. In Ref. \cite{coffman} it is defined through the density matrix of two qubits A and B, $\rho_{AB}$, in the following way. First define the "spin-flipped" density matrix $\tilde{\rho}_{AB} \equiv \left( \sigma_{2} \otimes \sigma_2 \right) \rho_{AB} \left( \sigma_{2} \otimes \sigma_2 \right)$

\begin{eqnarray}
\label{sigma2}
\sigma_2 = \begin{bmatrix}
    0       & -i \\
    i       & 0 
\end{bmatrix} \, ,
\end{eqnarray}

\noindent
where $\sigma_2$ is the Pauli matrix. Then, find the the eigenvalues $\lambda_i$ of the operator $\rho_{AB}\tilde{\rho}_{AB} $. Finally, the concurrence can be written as 

\begin{equation}
\label{conc2q}
 C = max \left\{ \lambda_1 - \lambda_2 - \lambda_3 - \lambda_4, 0 \right\}   ,
\end{equation}

\noindent
where the eigenvalues $\lambda_i$ are sorted in decreasing order.

Let us use the mutual information $I$ to measure the correlations between the two qubits. The mutual information measures the difference between the correlated state $\rho_{AB}$ and the uncorrelated state $\rho_A \otimes \rho_B$ and it is presented as

\begin{equation}
\label{mutinfo2q}
 I \left( \rho_{AB} \right) = S \left( \rho_{A} \right) + S \left( \rho_{B} \right) - S \left( \rho_{AB} \right)    ,
\end{equation}

\noindent
where $S \left( \rho \right) = -  \text{Tr} \left( \rho  \text{log}  \rho \right)$ is the Von Neumann entropy.
The mutual information, however, is unable to distinguish classical and quantum correlations and we use it as a check for the validity of the results given by the concurrence. Whenever the system has non-zero concurrence, the mutual information should also be non-zero, while the reverse statement does not hold. Moreover, the mutual information has no meaning for the case of mixed state and can only be used for pure states.

To deal with mixed states, let us turn to the negativity \textit{N}, which measures the entanglement of each qubit with the rest of the system. The negativity makes use of the positive partial transpose (PPT) criterion to quantify the entanglement in a system. The PPT criterion, first presented in Ref. \cite{peres}, says that if a state is separable, its density matrix has only positive eigenvalues. For the case of two qubits it represents a necessary and sufficient condition for the separability of a state. In general, density matrices $\rho$ have all positive eigenvalues and $\text{Tr} \rho = 1$. However, the partial transpose of a density matrix with respect to its subsystem A, denoted by $\rho^{T_A}$, might have some negative eigenvalues, while still maintaining $\text{Tr} \left( \rho^{T_A} \right) = 1$. Since separable states remain separable under partial transposition, if a partially transposed density matrix $\rho^{T_A}$ fails to have all positive eigenvalues, it means that the density matrix $\rho$ describes an entangled state. The negativity was defined by Vidal and Werner \cite{vidal} as

\begin{equation}
\label{neg}
 \textit{N}_A \left( \rho \right) = \frac{{\Vert \rho^{T_A} \Vert}_1 - 1 }{2}  ,
\end{equation}

\noindent
where $\Vert A \Vert_1 \equiv \text{Tr} \sqrt{A^{\dagger}A}$ is the trace norm.  An alternative way to calculate the negativity is to take the absolute value of the sum of the negative eigenvalues of the partial transpose density matrix of the system, which one can write as

\begin{equation}
\label{neg_abs}
 \textit{N}_A \left( \rho \right) = \frac{\sum_i \vert \lambda_i \vert - \lambda_i }{2}  ,
\end{equation}

\noindent
where $\lambda_i$ are all the eigenvalues of $\rho^{T_A}$. For separable states, whose density matrix only have positive eigenvalues, the negativity is zero. Thus, one can use the negativity to measure how much entanglement exists between the subsystem considered and the rest of the system, independently of its size.

\subsection{Quantum entanglement of three qubits}
For the case of three qubits, the quantum entanglement may arise in two ways. First, any pair of qubits can be entangled. Second, all qubits can be simultaneously entangled with each other. We use the negativity as a measure of the entanglement of one qubit with the rest of the system, therefore allowing us to detect entanglement between pairs. However, issues arise for the case of a system of more than two qubits. In particular, for this system the PPT criterion is only a necessary condition for separability, meaning that there can be entangled state even if $\textit{N}=0$. Nonetheless, if the negativity is found to be positive, then entanglement is present in the system.

The amount of entanglement that a qubit can share with a second qubit cannot be the same as the amount it shares with another one \cite{coffman}. This property of quantum entanglement is called the monogamy of entanglement, and it is one of its most fundamental properties. A monogamy relation for the three qubits case was explicitely found in Ref. \cite{coffman} and a quantity, called three-tangle, which quantifies the simultaneous entanglement of all three qubits was introduced. To detect the presence of GHZ states in the system, we need to measure the simultaneous entanglement of all the qubits. For this task, we make use of the three-$\pi$, which was introduced in Ref. \cite{ou}. The three-$\pi$ is defined in the same way as three-tangle \cite{coffman}, with the concurrence replaced by the negativity. In Ref. \cite{ou} a strong monogamy inequality was proven for the negativity, which allowed to introduce the three-$\pi$ as a measure of the simultaneous entanglement between three qubits where mixed states are considered.

First, one can find the residual entanglement for one of the three qubits A, B and C. For example, the residual entanglement for qubit A is

\begin{equation}
\label{resA}
\pi_A = \textit{N}_{A(BC)}^{2} - \textit{N}_{AB}^{2} - \textit{N}_{AC}^{2}.
\end{equation}

\noindent
However, this definition is not invariant under permutations of the qubits. Thus one needs to specify the residual entanglement for the other qubits (B and C) $\pi_B$ and $\pi_C$ to define a measure of entanglement which is invariant under permutations, the three-$\pi$

\begin{equation}
\label{3pi}
\pi_{ABC} = \frac{1}{3} \left( \pi_A + \pi_B + \pi_C  \right), 
\end{equation}

\noindent
as the average of all the residual entanglements.
The three-$\pi$ was proven to be a good measure of entanglement, satisfying the necessary conditions required in order to properly quantify entanglement listed in Ref. \cite{vedral}.
Since the three-$\pi$ is based on the negativity, it shares the same weaknesses. Namely, nonzero three-$\pi$ is only a necessary but not sufficient condition for the entanglement of the qubits and there can be entangled state with vanishing three-$\pi$.

\subsection{Quantum entanglement of $N$ qubits}
The $N$ qubit case has interesting applications to Josephson metamaterials. These systems, made from a collection of a large number of superconducting qubits, were used in experiments on the DCE \cite{lahte}. Due to quantum vacuum phenomena like the DCE and the DLE, quantum entanglement between the qubits of the Josephson metamaterial can arise. For the case of pure states of $N$ qubits, we propose to use another measure of entanglement defined in Ref. \cite{choi} as the square of convex-roof extended negativity (SCREN). As follows from Ref. \cite{choi}, a strong monogamy inequality holds for the SCREN, making it a good measure for the simultaneous entanglement of all the qubits in the system. However, the drawback with the SCREN is that it requires an optimization over all possible pure states decomposition of the system's density matrix in order to be used.

\section{System of N qubits coupled to a cavity}
\label{nq}
We give here an outline of the general method that can be used to treat a system with any number of qubits $N$. The Hamiltonian of the system is the Tavis-Cummings Hamiltonian \cite{tavis}

\begin{equation}
\label{H}
\hat{H}\left( t \right) = \hat{H}_{0} + \hat{H}_{I}\left( t \right) ,
\end{equation}

\noindent
where $\hat{H}_{0} $ is the unperturbed Hamiltonian and $\hat{H}_{I}\left( t \right) $ is the time-dependent interaction Hamiltonian. The unperturbed Hamiltonian reads

\begin{equation}
\label{H0}
\hat{H}_{0}  = \omega_{c} \hat{a}^{\dagger} \hat{a} + E_{0} \sum_{i=1}^{N} \hat{\sigma}_{i}^{+} \hat{\sigma}_{i}^{-} ,
\end{equation}

\noindent
where $\omega_{c}$ is the frequency of the cavity photons, $E_0$ is the transition frequency of the qubits, $\hat{a}$, $\hat{a}^{\dagger}$ are the creation and annihilation operators for the cavity photons and $\hat{\sigma}^{-} = \frac{\hat{\sigma}_1 -i \hat{\sigma}_2}{2}$, $\hat{\sigma}^{+} = \frac{\hat{\sigma}_1 +i \hat{\sigma}_2}{2}$ are defined through the Pauli matrices $\hat{\sigma}_1$ and $\hat{\sigma}_2$ for each qubit as the desctruction and creation operator of qubit excitations.

The interaction Hamiltonian is

\begin{equation}
\label{HI}
 \hat{H}_{I}\left( t \right) =  g \left( t \right) \left( \hat{a} + \hat{a}^{\dagger} \right) \sum_{i=1}^{N} \left( \hat{\sigma}_{i}^{-} + \hat{\sigma}_{i}^{+} \right) ,
\end{equation}

\noindent
where $g \left( t \right)$ is the time-dependent qubit/cavity coupling. As mentioned earlier, in order to give rise to the DLE, the switching of the qubit/cavity coupling must be done instantaneously. Furthermore, in Refs. \cite{shapiro, zhukov} it was found that  the periodic switching of the coupling, rather than single switching, greatly increases the DLE. For these reasons, the qubit/cavity coupling $g \left( t \right)$ is taken as

\begin{equation}
\label{g}
g \left( t \right) = g_0 \theta \left( \cos \varpi_s t \right),
\end{equation}

\noindent
where $g_0$ is the qubit/cavity coupling strength, $\theta \left( \cdot \right)$ is the Heaviside function that switches on periodically at a specified time and stays switched on during the period $T_s=1/\varpi_s$, where $\varpi_s$ is the frequency of the switching of the coupling. The possibility of turning on/off the qubit/cavity coupling was experimentally demonstrated in Refs. \cite{chen, lu}. It can be achieved by modulating the magnetic flux passing through an auxillary SQUID coupled to the qubit.

\subsection{Dynamical Lamb effect without dissipation: perturbative analytical approach}

As a first step, let us consider the non-dissipative system where the periodic switching of the qubit/cavity coupling is approximated by a constant value of the coupling after a single switching at time $t=0$. This approximation is valid if the frequency of switching of the coupling $\varpi_s \ll \omega_c + E_0$. Then, following Ref. \cite{remizov}, one can find an approximate solution for the time evolution of the system by solving the time-dependent Schroedinger equation perturbatively. That is, solving the time-dependent Schroedinger equation order by order for the perturbative expansion of the wavefunction $ \lvert \psi \left( t \right)  \rangle $ and the approximated Hamiltonian 

\begin{equation}
\label{H_t}
\hat{H}\left( t \right)  = \begin{cases} \hat{H}_0 , & \mbox{if } t=0 \\ \hat{H}_{avg}, & \mbox{if } t > 0 \end{cases},
\end{equation}

\noindent 
where $\hat{H}_{avg}$ is the time-average of Hamiltonian (\ref{H}) after the qubit/cavity coupling has been suddenly switched on at $t=0$

\begin{equation}
\label{Havg}
\hat{H}_{avg} \equiv \langle \hat{H}\left( t \right)  \rangle_t =  \hat{H}_{0} + \langle g \left( t \right) \hat{H}_{I} \left( t \right) \rangle_t ,
\end{equation}

\noindent
where $\langle \cdot  \rangle_t$ denotes time averaging and 

\begin{equation}
\label{HIavg}
\langle g \left( t \right) \hat{H}_{I}\left( t \right)  \rangle_t =  \langle g \left( t \right)  \left( \hat{a} + \hat{a}^{\dagger} \right) \sum_{i=1}^{N} \left( \hat{\sigma}_{i}^{-} + \hat{\sigma}_{i}^{+} \right) \rangle_t .
\end{equation}

\noindent
In the Schroedinger picture, the operators do not depend on time, therefore Eq. (\ref{HIavg}) becomes

\begin{equation}
\label{HIavg_g}
\langle g \left( t \right) \hat{H}_{I}\left( t \right)  \rangle_t =  \langle g \left( t \right) \rangle_t   \left( \hat{a} + \hat{a}^{\dagger} \right) \sum_{i=1}^{N} \left( \hat{\sigma}_{i}^{-} + \hat{\sigma}_{i}^{+} \right) ,
\end{equation}

\noindent
with 

\begin{eqnarray}
\label{g_avg}
\langle g \left( t \right) \rangle_t = \frac{1}{T} \int_{0}^{T} g \left( t \right) =   \frac{1}{T} \int_{0}^{T} g_0 \theta \left( \cos \varpi_s t \right) = \frac{g_0}{2},  
\end{eqnarray}

\noindent
where the coupling is averaged over the period of switching of the coupling $T=\frac{\pi}{\varpi_s}$, giving 

\begin{equation}
\label{HIavg_g0}
\langle g \left( t \right) \hat{H}_{I}\left( t \right)  \rangle_t =  \frac{g_0}{2}  \left( \hat{a} + \hat{a}^{\dagger} \right) \sum_{i=1}^{N} \left( \hat{\sigma}_{i}^{-} + \hat{\sigma}_{i}^{+} \right) .
\end{equation}

\noindent
The DLE arises because of the sudden switching of the qubit/cavity coupling at $t=0$ and the Hamiltonian (\ref{Havg}) gives an approximate description of the system under time periodical modulation of the qubit/cavity coupling for $t>0$.

In our approach, in order to perturbatively solve the time-dependent Schroedinger equation we consider the Hamiltonian (\ref{Havg}) in the Schroedinger picture

\begin{equation}
\label{schroedinger}
i \frac{d {\lvert \psi \left( t \right)  \rangle} }{dt} = \hat{H}_{avg} {\lvert \psi \left( t \right)  \rangle} .
\end{equation}

\noindent
We also truncate the infinite tower of possible photon states at a certain value of the photon number. Thus, the wavefunction for the system of $N$ qubits and $n$ photons can be written as

\begin{equation}
\label{psi}
\lvert \psi \left( t \right)  \rangle  = \sum_{i=0}^{n} \alpha_{gg...g, i} \left( t \right) \lvert gg...g , i   \rangle +  \alpha_{ge...g , i} \left( t \right) \lvert ge...g , i   \rangle + ... + \alpha_{ee...e , i } \left( t \right) \lvert ee...e , i   \rangle ,
\end{equation}

\noindent
where indices $g$ and $e$ correspond to ground and excited state of the qubit and $i$ counts the number of photons. Furthermore, if the interaction term in the Hamiltonian is small compared to the energy difference between the eigenvalues of the unperturbed Hamiltonian, one can perturbatively expand the wavefunction and the Hamiltonian in terms of the coupling strength $g_0/2$ and solve the time-dependent Schroendinger equation order by order:

\begin{equation}
\label{psi_exp}
{\lvert \psi \left( t \right)  \rangle} = {\lvert \psi \left( t \right)  \rangle}^{(0)} + \frac{g_0}{2} {\lvert \psi \left( t \right)  \rangle}^{(1)} +  \frac{g_0^2}{4} {\lvert \psi \left( t \right)  \rangle}^{(2)} + ... \, ,
\end{equation}

\begin{equation}
\label{H_exp}
 \hat{H} =\hat{H}_{0} + \frac{g_0}{2} \langle \hat{H}_{I}\left( t \right)  \rangle_t,
\end{equation}

\noindent
where the small parameter is the qubit/cavity coupling $\left( g_0/2 \right)^n $. As a result one obtains the following set of differential equations. At zero-th order:
 
 \begin{equation}
\label{schroedinger_0}
i \frac{d {\lvert \psi \left( t \right)  \rangle}^{(0)} }{dt} = \hat{H}_0  {\lvert \psi \left( t \right)  \rangle}^{(0)} .
\end{equation}

\noindent
The latter allows to obtain the coefficients' equations (seeking the simplicity in notation, here and below we are omitting the argument "$t$"  for the time-dependent coefficients $\alpha \left( t \right)$ )

\begin{equation}
\label{alpha_0}
i \frac{d {\alpha}_{\bar{x}, n }^{(0)} }{dt} = \left[ \omega_c n + E_{0} \left( \bar{x} \cdot \bar{1} \right) \right] {\alpha}_{\bar{x},n }^{(0)} ,
\end{equation}

\noindent
where $\bar{x}$ stands for the $N$-bit string which represents the state of the qubits as a string of zeros (for the ground state $g$) and ones (for the excited state $e$), e.g. $\bar{x} \equiv 001...1 = gge...e$. Also, $\bar{x} \cdot \bar{1}$ is the dot product between the $N$-bit string and the string of all ones, namely $\bar{x} \cdot \bar{1} = x_0 1 + x_1 1 + x_2 1 + ... + x_N 1$, which counts the number of qubit's excitations in the system.
The general recursive differential equation for any other order $(j)$ has the following form

\begin{equation}
\label{schroedinger_j}
i \frac{d {\lvert \psi \left( t \right)  \rangle}^{(j)} }{dt} = \hat{H}_0  {\lvert \psi \left( t \right)  \rangle}^{(j)} +  \langle \hat{H}_{I}\left( t \right)  \rangle_t  {\lvert \psi \left( t \right)  \rangle}^{(j-1)} ,
\end{equation}

\noindent
that can be reduced to the set of differential equations for the coefficients

\begin{equation}
\begin{split}
\label{alpha_j}
i \frac{d {\alpha}_{x_0x_1 \ldots x_N,n }^{(j)} }{dt} = \left[ \omega_c n + E_{0} \left( \bar{x} \cdot \bar{1} \right) \right] {\alpha}_{x_0x_1 \ldots x_N,n }^{(j)} +  \sum_{i=0}^{N} \left( \sqrt{n} \delta_{ x_{i} -1, 0} {\alpha}_{x_0x_1 \ldots x_{i} -1 \dots x_N,n-1 }^{(j-1)} + \right. \\
\left. + \sqrt{n+1} \delta_{ x_{i} +1, 1} {\alpha}_{x_0x_1 \ldots x_{i} +1 \dots x_N,n+1 }^{(j-1)} +  \sqrt{n} \delta_{ x_{i} +1, 0} {\alpha}_{x_0x_1 \ldots x_{i} +1 \dots x_N,n-1 }^{(j-1)} + \sqrt{n+1} \delta_{ x_{i} -1, 1} {\alpha}_{x_0x_1 \ldots x_{i} -1 \dots x_N,n+1 }^{(j-1)} \right) .
\end{split}
\end{equation}

\noindent
In Eq. (\ref{alpha_j}), $x_{i}$ denotes the $i$-th element of the $N$-bit string $\bar{x}$ in the 0,1 notation. Therefore, Eq. (\ref{alpha_j}) gives the differential equation for any coefficient of the $N$-bit state specified by $\bar{x}, n$, at any order $j$.
Now, solving for a certain initial value $ \alpha_{\bar{x}} \left( 0 \right) $, one can find the time evolution of the coefficients $\alpha \left( t \right)$ and thus from Eqs. (\ref{psi}) and (\ref{psi_exp}) the wavefunction. In particular, one has to solve a system of $2^N \times (n+1)$ differential equations for the coefficients $\alpha \left( t \right)$.

These perturbative analytical solutions of Eq. (\ref{schroedinger}) are just an approximation of the ones for the system considered, valid for low frequency of switching of the coupling $\varpi_s \ll \omega_c + E_0$. To have a more accurate description of the real system, one must consider fast periodic modulations of the coupling strength which are needed to amplify the effect of the DLE. We are interested in finding these solutions to provide a check to the numerical procedure presented in the following Sections. 


\subsection{Dynamical Lamb effect with dissipation: numerical approach}

Dissipation, that is a result of losses in the system, is an important factor in the description of the time evolution of a quantum system which interacts with the environment. In the previous work \cite{berman, amico} on the quantum entanglement of two and three qubits coupled to a nonstationary cavity this was not taken into account. We consider the effects of dissipation by taking the system to be weakly coupled to a memory-less reservoir. By considering the system-reservoir coupling to be weak, we are assuming that the reservoir has a neglegible influence on the system (Born-Oppenheimer approximation \cite{born}). Furthermore, a memory-less reservoir is a reservoir whose correlations with the system decay much faster than the relaxation time of the system itself. Thus, the reservoir does not have a memory of previous states of the system (Markov approximation). One can describe the non-unitary dynamics of the system through the Lindblad master equation \cite{gorini, lindblad} for the system's density matrix $\rho_{s} \left( t \right)= \text{Tr}_{env} \left[ \rho \left( t \right) \right]$ with Hamiltonian $\hat{H}\left( t \right)$

\begin{equation}
\label{lindblad}
\frac{d \rho_{s} \left( t \right)  }{dt} = -i \left[ \hat{H}\left( t \right) , \rho_{s} \left( t \right) \right] + \frac{\gamma_j}{2} \sum_{j} 2 \hat{A}_j      \rho_{s} \left( t \right) \hat{A}_j^{\dagger} -  \rho_{s} \left( t \right) \hat{A}_j^{\dagger} \hat{A}_j  -  \hat{A}_j^{\dagger} \hat{A}_j \rho_{s} \left( t \right) . 
\end{equation}

\noindent
In Eq. (\ref{lindblad}), $\hat{H}\left( t \right)$ is the Hamiltonian (\ref{H0}) of the qubit/cavity subsystem excited by the DLE, $\hat{A}_j$ is the $j^{th}$ subsystem's annihilation operator, $\hat{A}_j^{\dagger}$ its creation operator and  $\gamma_j$ is the corresponding decay rate. In our case, we take $\hat{A}_j = \hat{a}, \hat{\sigma}^{-}_{1}, \hat{\sigma}^{-}_{2}, ..., \hat{\sigma}^{-}_{N},\hat{\sigma}^{(3)}_{1}, \hat{\sigma}^{(3)}_{2}, ..., \hat{\sigma}^{(3)}_{N}$, thus accounting for the cavity and the qubits interaction with the environment which causes excitation/relaxation in the system and qubit dephasing. The decay rates are indicated as $k$ and $\gamma_1, \gamma_2, ..., \gamma_N$ for the cavity photons and the qubits, respectively, while $\gamma_{\phi_1}, \gamma_{\phi_2}, ..., \gamma_{\phi_N}$ denote the qubit's dephasing rate. For superconducting qubits coupled to a coplanar waveguide, playing the role of the cavity, the dominant source of relaxation is the Purcell effect \cite{purcell, tureci}. The latter is the increase/decrease in the decay rate $ \gamma$ of the qubit when its transition frequency is in-resonance/off-resonance with the frequency of the cavity mode. The qubit decay rate due to the Purcell effect is given by $\gamma \approx \kappa\frac{\lambda^2}{\left( \omega_c - E_0 \right)^2 }$. The decay rate of the qubit is related to its relaxation time as ${T_1}_q = \frac{1}{\gamma}$. For the transmon superconducting qubit \cite{transmon}, the dephasing time ${T_2}_q = \frac{1}{\gamma_{\phi} }$, which is the time it takes to lose information about the qubit's phase, is limited by the relaxation time ($T_2 \sim 2 T_1$).  For the resonant mode of the cavity, it was experimentally shown in Refs. \cite{burnett,gao,pappas} that at low temperature the main source of dissipation comes from its coupling with parasitic two level systems present at the cavity/substrate interface. The decay rate of the cavity mode $\kappa$ can be found from the lifetime $T_{ph}$ of the photons in the cavity as $ \kappa = \frac{1}{T_{ph}}$. In Ref. \cite{gao}, the lifetime of the cavity photons was estimated by studying their interaction with two level systems and the validity of the model was confirmed through experiments. Furthermore, in Ref. \cite{bruno} it was found that the relaxation time of the resonant mode of the cavity can be greatly improved by careful engineering of the fabrication techniques of the cavity.
Since the main sources of dissipation for the qubits and the photons are unrelated, we model the two different relaxation channels as separate environments for the qubit and the cavity mode. This means that we consider the qubit as interacting with a bath which has certain parameters and the cavity mode as interacting with another independent bath characterized by different parameters. As a result, we can apply Eq. (\ref{lindblad}) specifying the system's Hamiltonian and the subsystem's annihilation operators $\hat{A}_j$ for the case at hand.
Since the number of equations describing the time-evolution of the density matrix elements grows exponentially with the number of qubits in the cavity, an analytical solution of the problem is not viable. Thus, the master equation is solved numerically by using the QuTip software \cite{qutip, qutip2}.

\section{Two qubits and a cavity mode}
\label{2q}
Let us first consider the case of two qubits coupled to the same cavity. The Tavis-Cummings Hamiltonian for the case $N=2$, is

\begin{equation}
\label{H2q}
\hat{H}  = \omega_{c} \hat{a}^{\dagger} \hat{a} + E_{0}^{(1)} \hat{\sigma}_{1}^{+} \hat{\sigma}_{1}^{-} + E_{0}^{(2)} \hat{\sigma}_{2}^{+} \hat{\sigma}_{2}^{-} + g_{1} \left( t \right) \left( \hat{a} + \hat{a}^{\dagger} \right) \left( \hat{\sigma}_{1}^{-} + \hat{\sigma}_{1}^{+} \right) + g_{2} \left( t \right) \left( \hat{a} + \hat{a}^{\dagger} \right) \left( \hat{\sigma}_{2}^{-} + \hat{\sigma}_{2}^{+} \right),
\end{equation}

\noindent
where the index 1 and 2 is used to refer to operators or quantities relative to the first and second qubits respectively. To achieve nonadiabatic modulation of the qubit/cavity coupling, we assume a square-wave time-dependent coupling as in Eq. (\ref{g}).
Following what was done in the previous Section, we first develop an analytical perturbative treatment for the case of two qubits coupled to a cavity with constant coupling in the absence of losses. Then we consider the case of two qubits coupled to a cavity with the qubit/cavity coupling periodically switched on/off nonadiabatically in presence of dissipation. The Lindblad equation for the system interacting with a dissipative environment is numerically solved. The results obtained from the analytical and numerical calculations are shown in Figs. \ref{num_an_2q} - \ref{span_k_2q} and discussed in Sec. \ref{results}.

\subsection{Dynamical Lamb effect without dissipation: perturbative analytical and numerical calculations}

As described in Sec. \ref{nq}, for a system with no dissipation and constant qubit/cavity coupling after being switched on at $t=0$, one can solve the time-dependent Schroedinger equation (\ref{schroedinger}) perturbatively order by order like it was done in Ref. \cite{remizov}. By doing so, one finds a set of differential equations for the time-dependent coefficients $\alpha (t)$ of the wavefunction. Solving the system of equations yields the time evolution of the wavefunction for a fixed order of the perturbation. We emphasize that the purpose of the analytical derivation is to provide a frame of reference for the the numerical solutions of the Schroedinger equation. Comparing the agreement between the perturbative calculations and the numerical ones, gives an indication of the correctness of the numerical calculations.
For the case of two qubits, it follows from Eq. (\ref{psi}) that the wavefunction takes the form

\begin{equation}
\label{psi2q}
\lvert \psi \left( t \right)  \rangle  = \sum_{i=0}^{n} \alpha_{gg, i} \left( t \right) \lvert gg , i   \rangle +  \alpha_{ge, i} \left( t \right) \lvert ge , i   \rangle + \alpha_{eg , i } \left( t \right) \lvert eg , i   \rangle + \alpha_{ee, i } \left( t \right) \lvert ee, i   \rangle .
\end{equation}

\noindent
If the interaction Hamiltonian $ \hat{H}_{I}$ can be considered a small correction to the non-interacting Hamiltonian $\hat{H}_0$, one can expand the wavefunction and the Hamiltonian as shown in Eqs. (\ref{psi_exp}) and (\ref{H_exp}), respectively, with the non-interacting Hamiltonian $\hat{H}_0 = \omega_c \hat{a}^{\dagger}\hat{a} + E_0 \hat{\sigma}^{+}_{1}\hat{\sigma}^{-}_{1} + E_0 \hat{\sigma}^{+}_{2}\hat{\sigma}^{-}_{2}$ and $ \langle g \left( t \right) \hat{H}_{I}  \rangle_t = \frac{g_0}{2} \left( \hat{a}^{\dagger} + \hat{a} \right) \left( \hat{\sigma}^{+}_{1} + \hat{\sigma}^{-}_{1} + \hat{\sigma}^{+}_{2} + \hat{\sigma}^{-}_{2} \right)  $.

One can then solve the Schroedinger equation order by order in the parameter $\frac{g_0}{2}$ as shown in the previous Section.
In Appendix \ref{app_2q}, the wavefunction (\ref{psi2q}) is perturbatively expanded up to second order in terms of $\frac{g_0}{2}$, obtaining the set of differential equations (\ref{alpha_j}) for the time-dependent coefficients $\alpha \left( t \right)$. Then, truncating the perturbative expansion (\ref{psi_exp}) of the wavefunction (\ref{psi2q}) at second order and considering only one photon in the cavity, we obtain the following approximate solution of the Schroedinger equation

\begin{eqnarray}
\label{psi_exp2q2}
\begin{split}
{\lvert \psi \left( t \right)  \rangle} =  {\lvert gg,0  \rangle}^{(0)} + \left\{  \frac{g_0}{2} \frac{1}{\omega + E_0} \left( e^{-i \left( \omega + E_0 \right) t} -1 \right) \left[ {\lvert ge,1  \rangle}^{(1)} + {\lvert eg,1  \rangle}^{(1)} \right] \right\} + \\ 
+ \left\{ \frac{g_0^2}{2} \frac{1}{\left( \omega + E_0 \right)^2} \left( i \left( \omega + E_0 \right) t +  e^{-i \left( \omega + E_0 \right) t} -1 \right) {\lvert gg,0  \rangle}^{(2)} + \right. \\
 \left. + \frac{g_0^2}{4}\frac{1}{E_0 \left( \omega + E_0 \right)^2 \left( E_0 -  \omega  \right)} \left[ 2E_0 - 2E_0 e^{-i \left( \omega + E_0 \right) t} + \left( \omega + E_0 \right) \left( e^{-i \left(  2E_0 \right) t} -1 \right) \right] {\lvert ee,0  \rangle}^{(2)}      \right\},
\end{split}
\end{eqnarray}

\noindent
where we consider the system to initially be in its ground state by imposing the initial condition ${\lvert \psi \left( 0 \right)  \rangle}^{(0)} = {\lvert gg,0  \rangle}^{(0)} $.

\begin{figure}[t]
\subfloat[]{\label{fig:concurrence}}{\includegraphics[width = 2 in, height = 1.5 in]{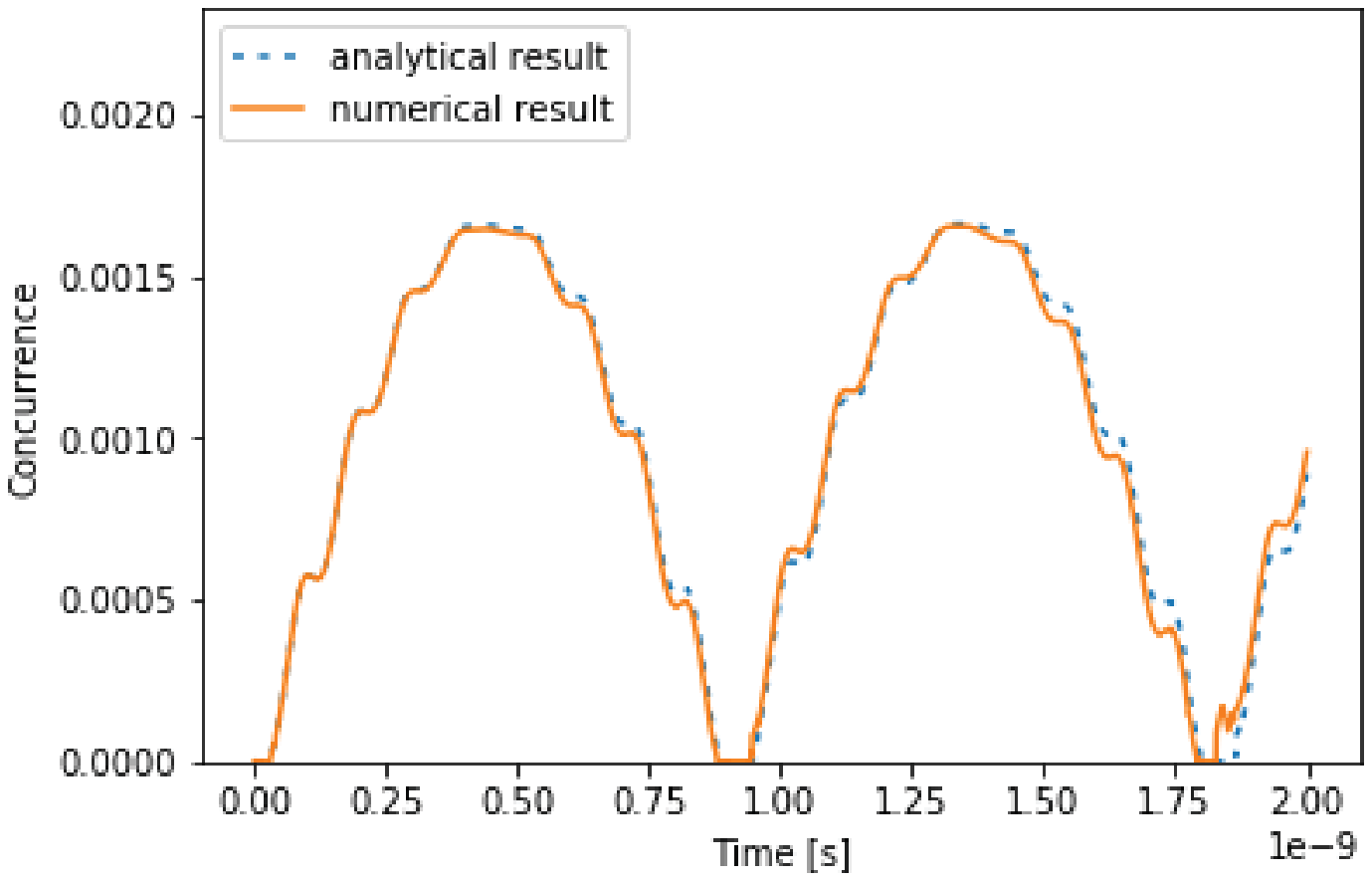}} 
\subfloat[]{\label{fig:mutual information}}{\includegraphics[width = 2in, height = 1.5 in]{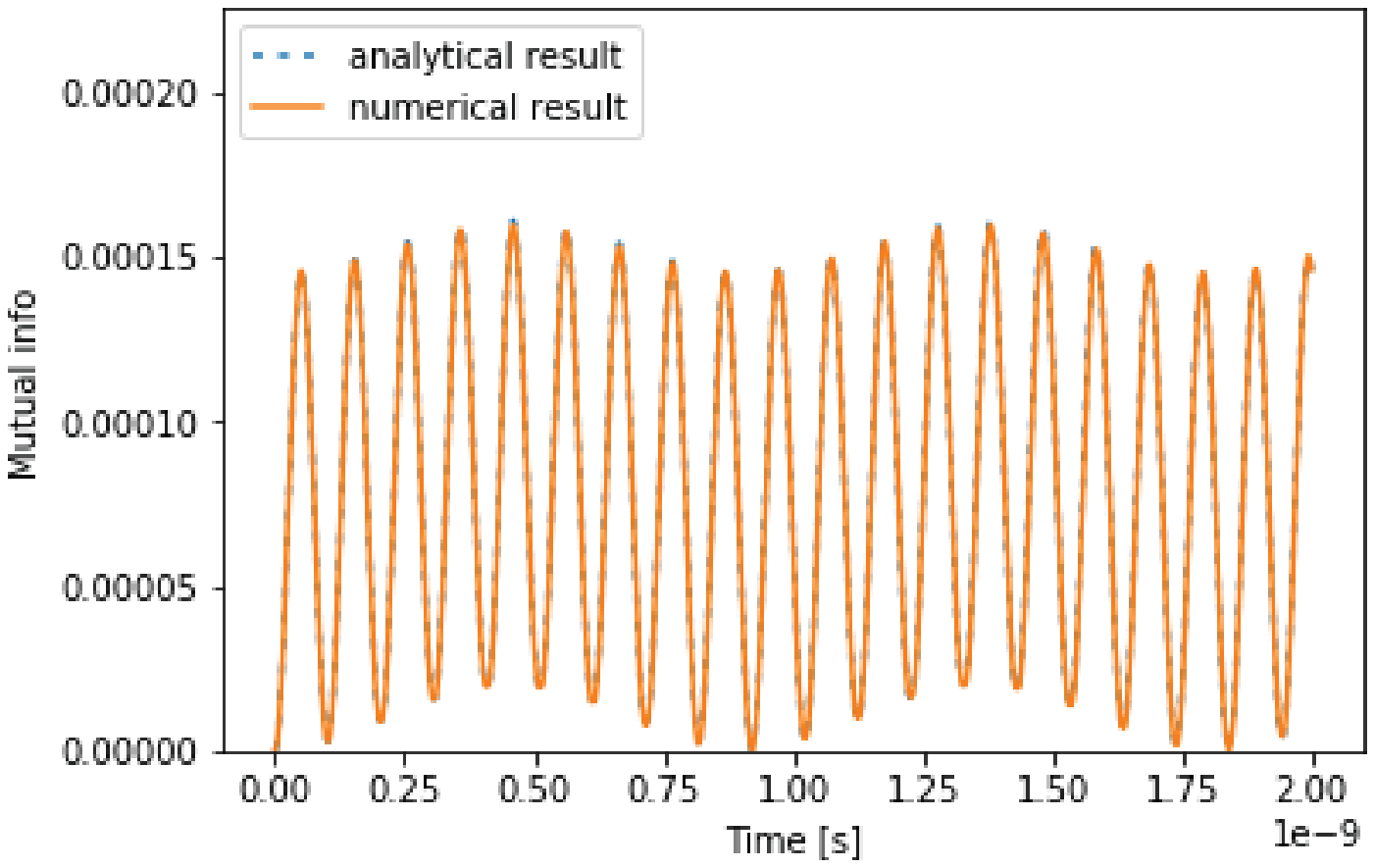}} 
\subfloat[]{\label{fig:negativity}}{\includegraphics[width = 2in, height = 1.5 in]{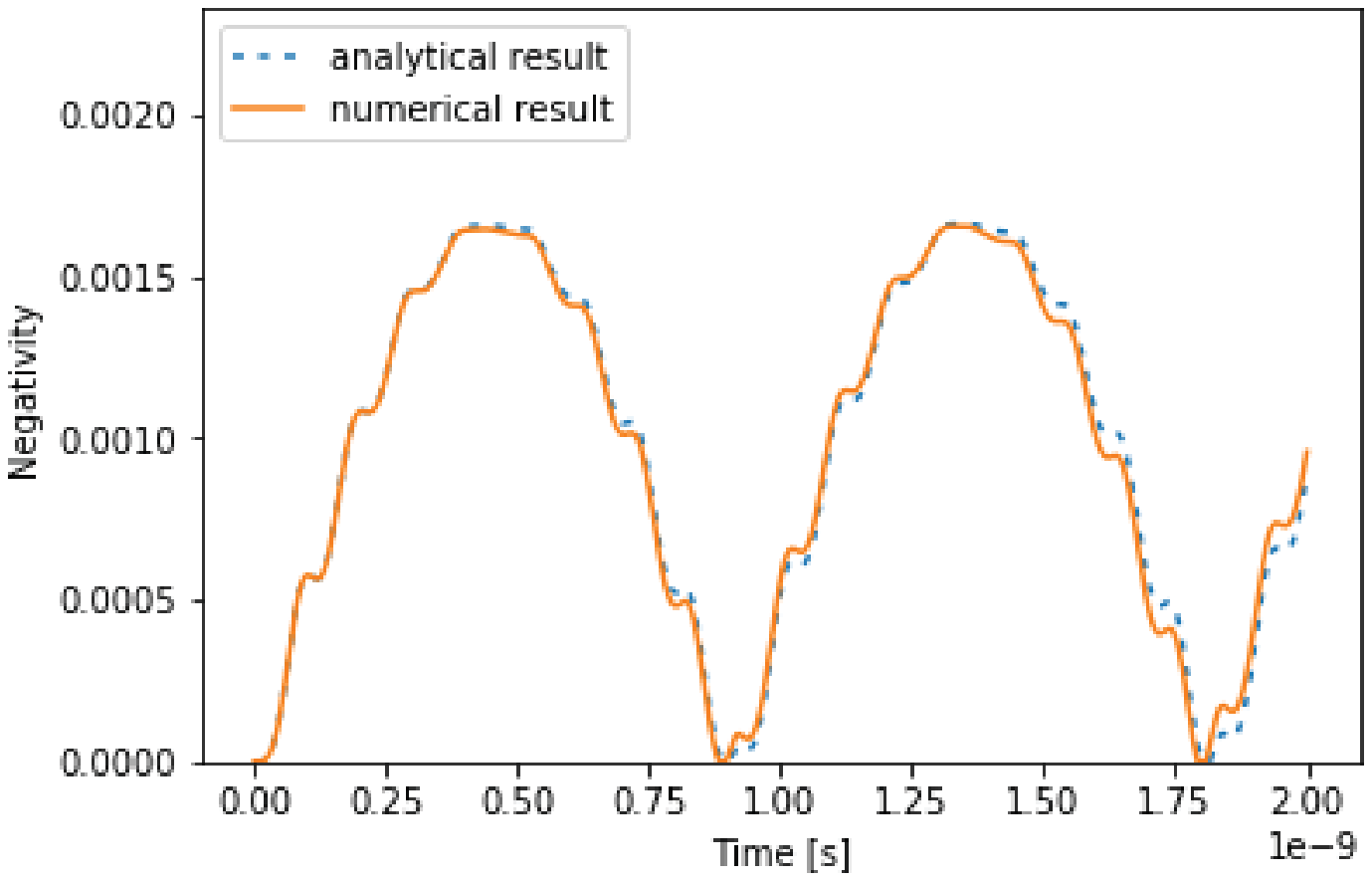}}
\caption{The comparison between analytical and numerical calculations for the time evolution of different measures of entanglement. (a) Concurrence, (b) mutual information, and (c) negativity.}
\label{num_an_2q}
\end{figure}

With this solution at hand, one can calculate the measures of entanglement presented in Sec. \ref{ment}. In particular, we use the wavefunction (\ref{psi_exp2q2}) derived with a perturbative analytical approach and substitute it into Eqs. (\ref{conc2q}), (\ref{mutinfo2q}), (\ref{neg_abs}) to calculate the concurrence, the mutual information and the negativity, respectively.
The time evolution of the system is also found by numerically solving the time-dependent Schroedinger equation (\ref{schroedinger}) using the same Hamiltonian and initial wavefunction. The values of parameters used in the calculations for the qubit and cavity frequencies and the qubit/cavity coupling strength are taken from Ref. \cite{lu} and are typical values for an experimental setup. Namely, $\omega_c = 2 \pi \times 4.343$ GHz, $E_0 = 2 \pi \times 5.439$ GHz, $\frac{g_0}{2} = 2 \pi \times 50$ MHz and $\varpi_s \ll 2 \pi \times 10.782$ GHz. The comparison between the results obtiained from the perturbative analytical and the numerical approaches is shown in Fig. \ref{num_an_2q}.

\subsection{Dynamical Lamb effect with dissipation: numerical calculations}

Let us now consider the case where the qubit/cavity coupling is periodically switched-on/off nonadiabatically in the presence of dissipation. The sudden switching of the coupling reproduces the conditions required for the dynamical Lamb effect to arise, leading to the excitation of the qubits and the creation of photons. Furthermore, the two qubits can be parametrically entangled using the DLE. To study the time evolution of the quantum entanglement between the qubits under the driving of the coupling, we numerically integrate Lindblad's master equation (\ref{lindblad}) which takes the form

\begin{equation}
\label{lindblad_2q}
\begin{split}
\frac{d \rho_{s} \left( t \right)  }{dt} = -i \left[ \hat{H}\left( t \right) , \rho_{s} \left( t \right) \right] 
+ \frac{k}{2} \left\{ 2 \hat{a}     \rho_{s} \left( t \right) \hat{a}^{\dagger} -  \rho_{s} \left( t \right) \hat{a}^{\dagger} \hat{a}  -  \hat{a}^{\dagger} \hat{a} \rho_{s} \left( t \right) \right\} +  \\
+ \sum_{i=1}^{2} \frac{\gamma_j}{2} \left\{ 2 \hat{\sigma}^{-}_{j}      \rho_{s} \left( t \right) \hat{\sigma}^{+}_{j} -  \rho_{s} \left( t \right) \hat{\sigma}^{+}_{j} \hat{\sigma}^{-}_{j} -  \hat{\sigma}^{+}_{j} \hat{\sigma}^{-}_{j} \rho_{s} \left( t \right) \right\} 
+ {\gamma_{\phi_j}} \left\{ \hat{\sigma}^{(3)}_{j}     \rho_{s} \left( t \right) \hat{\sigma}^{(3)}_{j} -  \rho_{s} \left( t \right)  \right\},
\end{split} 
\end{equation}

\noindent
where $\hat{H}\left( t \right)$ is the system's Hamiltonian (\ref{H2q}), $\hat{\sigma}^{\pm}_{j}$ are the creation and destruction operators for excitation of the $j$-th qubit, $ \hat{\sigma}^{(3)}_{j}$ is the Pauli matrix for the $j$-th qubit and $k, \gamma_j, \gamma_{\phi_j}$ take into account possible channels of dissipation of the cavity and the qubit in the form of qubit and cavity relaxation and qubit dephasing. In the numerical calculations, we use realistic values of the parameters of the system taken from the experiment done in \cite{lu}. Namely, $E_0^{(1)} = E_0^{(2)} \equiv E_0 =  2 \pi \times 5.439$ GHz for the transition frequencies of the qubits, $\omega_c = 2 \pi \times 4.343$ GHz for the frequency of the cavity photons, $g_1 \left( t \right) = g_2 \left( t \right) = g_0 \theta \left( \cos{\varpi_s t} \right)$ with $g_0 = 2 \pi \times 300$ MHz for the qubit/cavity coupling. $\kappa = 2 \pi \times 1.6$ MHz for the relaxation rate of the cavity, $\gamma_1 = \gamma_2 = 2 \pi \times 7.6$ MHz for the relaxation rate of the qubits and $\gamma_{\phi_1} = \gamma_{\phi_2} = 2 \pi \times 3$ MHz for the dephasing rate of the qubits.

\begin{figure}[h]
\subfloat[]{\label{fig:concurrence_dle}}{\includegraphics[width = 2 in, height = 1.5 in]{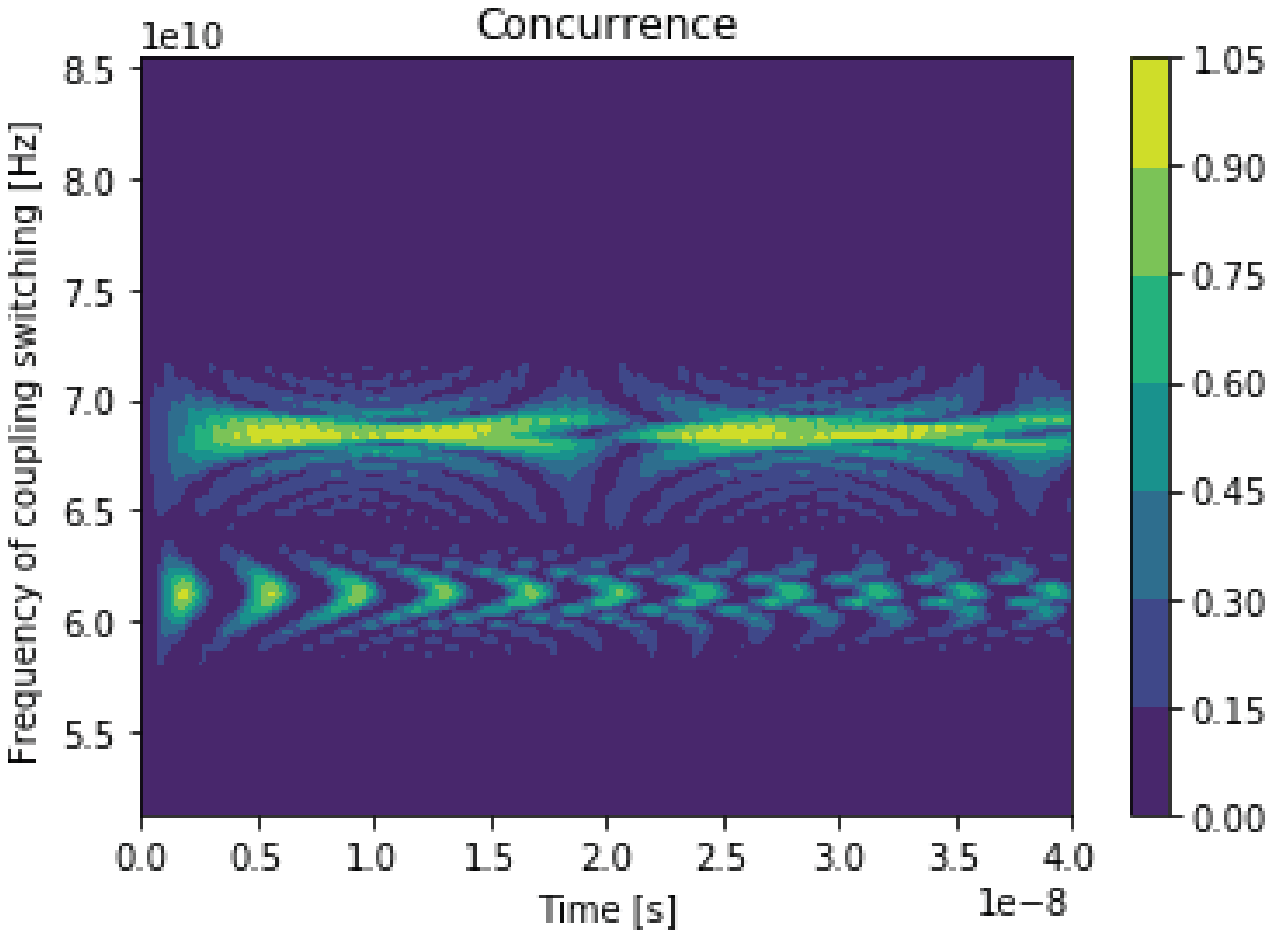}} 
\subfloat[]{\label{fig:negativity_dle}}{\includegraphics[width = 2in, height = 1.5 in]{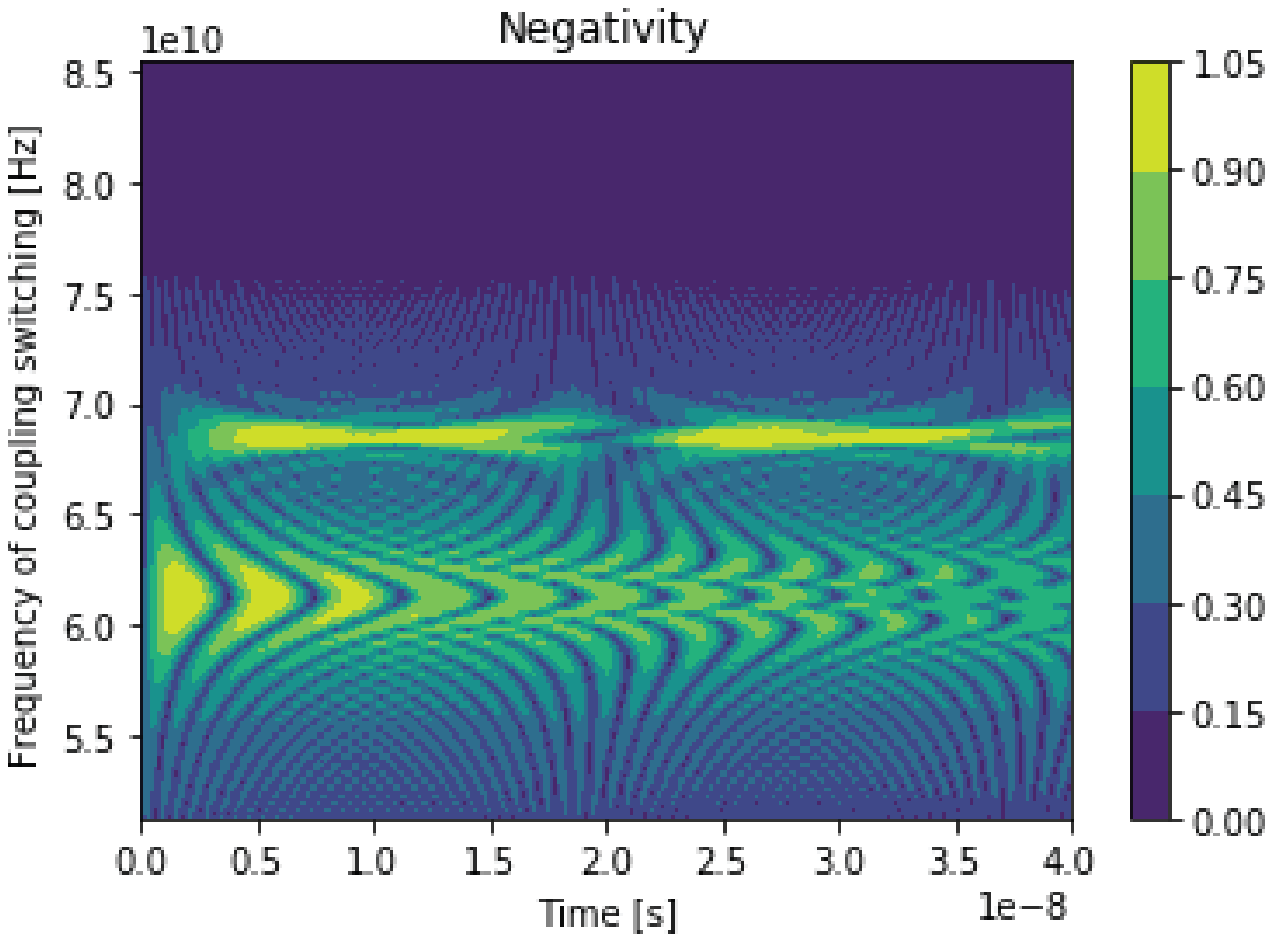}}
\caption{Time evolution of different measures of entanglement for a range of frequencies of switching of the coupling. (a) Concurrence and (b) negativity.}
\label{span_f_2q}
\end{figure}

\begin{figure}[h]
\subfloat[]{\label{fig:concurrence_dle_phot}}{\includegraphics[width = 2 in, height = 1.5 in]{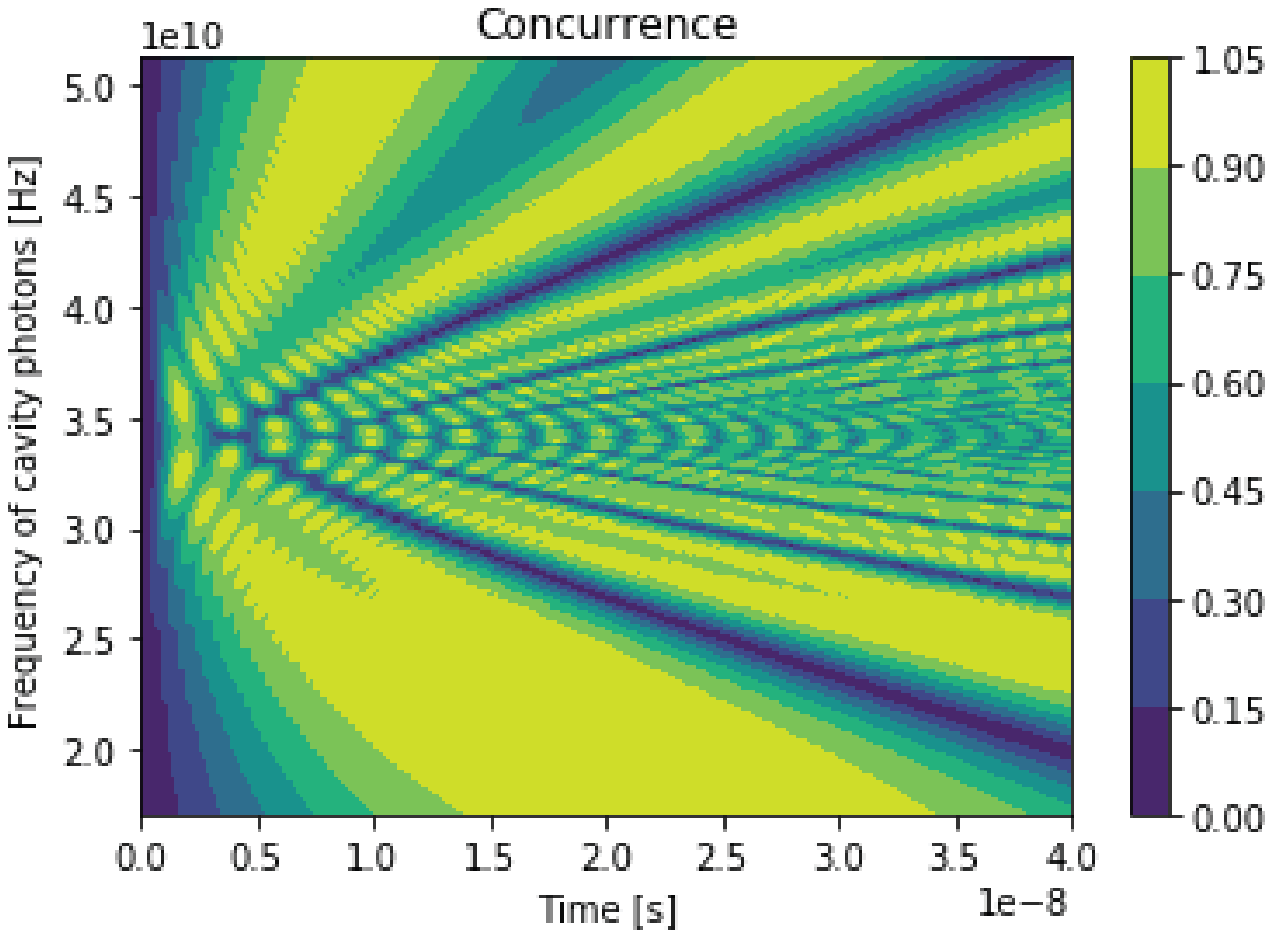}} 
\subfloat[]{\label{fig:negativity_dle_phot}}{\includegraphics[width = 2in, height = 1.5 in]{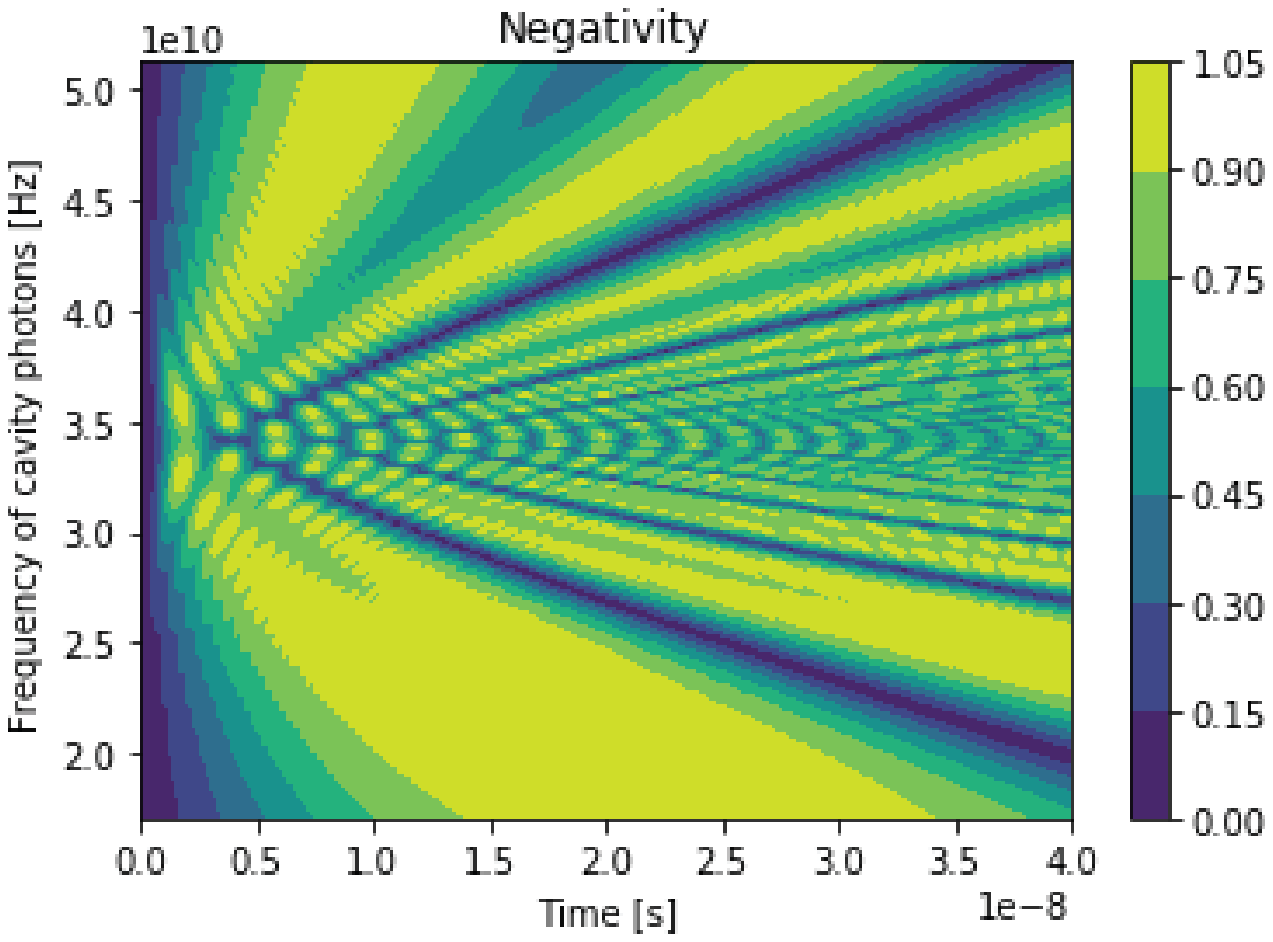}}
\caption{Time evolution of different measures of entanglement for a range of frequencies of cavity photons. (a) Concurrence and (b) negativity.}
\end{figure}

\begin{figure}[ht]
\subfloat[]{\label{fig:concurrence_dle_diss}}{\includegraphics[width = 2 in, height = 1.5 in]{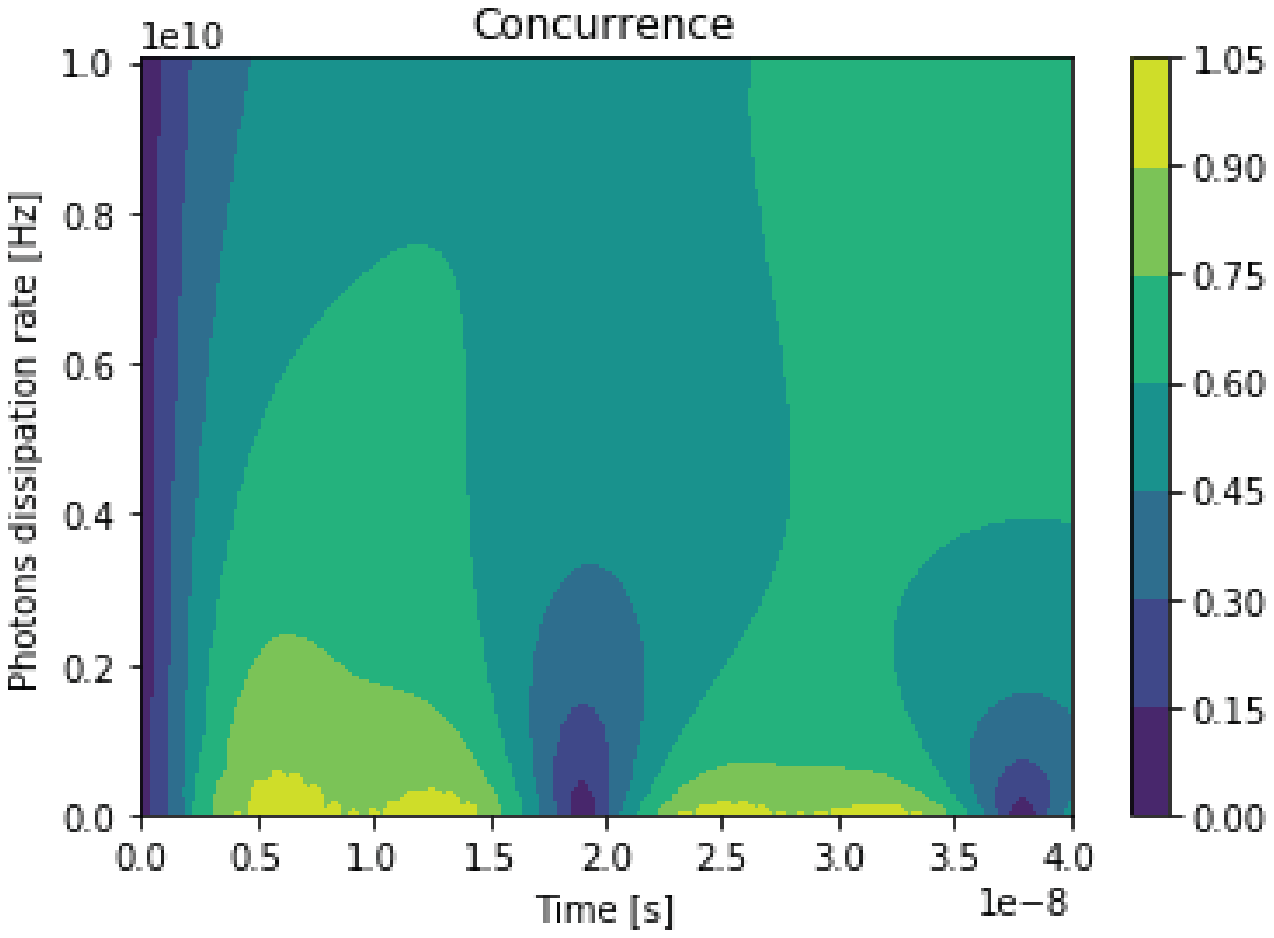}} 
\subfloat[]{\label{fig:negativity_dle_diss}}{\includegraphics[width = 2in, height = 1.5 in]{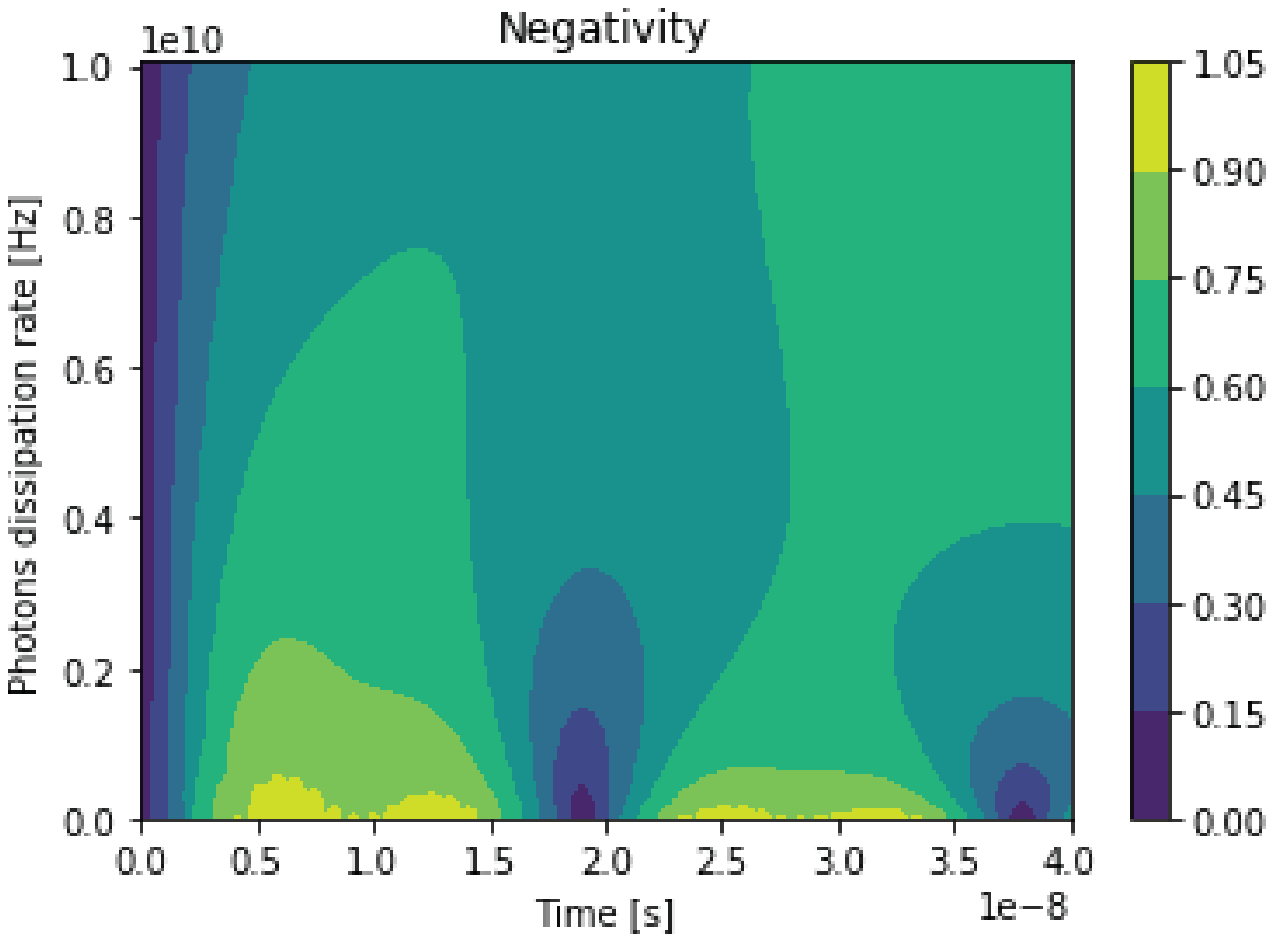}}
\caption{Time evolution of different measures of entanglement for a range of decay rates $\kappa$ of the cavity photons. (a) Concurrence and (b) negativity.}
\label{span_k_2q}
\end{figure}

The results of our calculations are presented in Figs. \ref{span_f_2q} - \ref{span_k_2q}. To measure the entanglement of the system, we rely on the concurrence and the negativity only, since the mutual information cannot be used for mixed states. The change in time dependence of the quantum entanglement between the qubits when the frequency of switching of the coupling $\varpi_s$ is tuned over the range $\varpi_s \in \left[ E_0 , 4E_0 \right]$ is depicted in Figs. \ref{fig:concurrence_dle} and \ref{fig:negativity_dle}.

In Figs. \ref{fig:concurrence_dle_phot}, \ref{fig:negativity_dle_phot} the time evolution of the quantum entanglement is studied when the frequency of the cavity photons is changed over the range $\omega_c \in \left[ \frac{E_0}{2},\frac{3 E_0}{2} \right] $. The frequency of switching of the coupling is set at the sum frequency of the transition frequencies of the qubits $\varpi_s = 2 E_0$. All the fixed parameters take the same values specified earlier.

The dependence of the quantum entanglement on the cavity dissipation rate $\kappa$ is studied and the results are presented in Figs. \ref{fig:concurrence_dle_diss} and \ref{fig:negativity_dle_diss} in the range $\kappa  \in \left[ 0, 2\pi \times1.6  \, \text{GHz} \right]$.

\section{Three qubits and a cavity mode}
\label{3q}
Let us now consider the case of three qubits coupled to a common cavity. The Hamiltonian can be obtained from Eq. (\ref{H}), specifying $N=3$

\begin{equation}
\begin{split}
\label{H3q}
\hat{H}  = \omega_{c} \hat{a}^{\dagger} \hat{a} + E_{0}^{(1)} \hat{\sigma}_{1}^{+} \hat{\sigma}_{1}^{-} + E_{0}^{(2)} \hat{\sigma}_{2}^{+} \hat{\sigma}_{2}^{-} + E_{0}^{(3)} \hat{\sigma}_{3}^{+} \hat{\sigma}_{3}^{-} + \\
+ g_{1} \left( t \right) \left( \hat{a} + \hat{a}^{\dagger} \right) \left( \hat{\sigma}_{1}^{-} + \hat{\sigma}_{1}^{+} \right) + g_{2} \left( t \right) \left( \hat{a} + \hat{a}^{\dagger} \right) \left( \hat{\sigma}_{2}^{-} + \hat{\sigma}_{2}^{+} \right) + g_{3} \left( t \right) \left( \hat{a} + \hat{a}^{\dagger} \right) \left( \hat{\sigma}_{3}^{-} + \hat{\sigma}_{3}^{+} \right) .
\end{split}
\end{equation}

Following what was done in the previous Section, we first develop an analytical perturbative treatment for the case of three qubits coupled to a cavity with constant coupling and in the absence of losses. This is used to compare the numerical solutions of the Schroedinger equation with the analytical ones. Then, we numerically solve the Lindblad equation describing three qubits coupled to a cavity where the qubit/cavity coupling is periodically switched on/off nonadiabatically in presence of dissipation. The results of the calculations are presented in Figs. \ref{num_an_3q} - \ref{ent2qub} and discussed in Sec. \ref{results}.

\subsection{Dynamical Lamb effect without dissipation: perturbative analytical and numerical calculations}

For a non-dissipative system where the qubit/cavity coupling is turned on at $t=0$, giving rise to the DLE, and then is fixed to a constant value, it is possible to find a simple perturbative solution of Eq. (\ref{schroedinger}) following the same procedure as in the previous Section. For the case of three qubits, from Eq. (\ref{psi}) we get the following wavefunction

\begin{eqnarray}
\label{psi3q}
\lvert \psi \left( t \right)  \rangle  = \sum_{i=0}^{n} \alpha_{ggg, i} \left( t \right) \lvert ggg , i   \rangle +  \alpha_{gge, i} \left( t \right) \lvert gge , i   \rangle + \alpha_{geg , i } \left( t \right) \lvert geg , i   \rangle + \alpha_{gge , i } \left( t \right) \lvert gge , i   \rangle + \nonumber \\
+ \alpha_{gee, i } \left( t \right) \lvert gee, i   \rangle + \alpha_{ege, i } \left( t \right) \lvert ege, i   \rangle + \alpha_{eeg, i } \left( t \right) \lvert eeg, i   \rangle + \alpha_{eee, i } \left( t \right) \lvert eee, i   \rangle .
\end{eqnarray}

We take the system to be initially in the ground state ${\lvert \psi \left( 0 \right)  \rangle}^{(0)} = {\lvert ggg,0  \rangle}^{(0)} $ . Taking the interaction Hamiltonian $ \langle g \left( t \right) \hat{H}_{I}  \rangle_t = \frac{g_0}{2} \left( \hat{a}^{\dagger} + \hat{a} \right) \left( \hat{\sigma}^{+}_{1} + \hat{\sigma}^{-}_{1} + \hat{\sigma}^{+}_{2} + \hat{\sigma}^{-}_{2} + \hat{\sigma}^{+}_{3} + \hat{\sigma}^{-}_{3} \right)  $ as a small correction of the non-interacting Hamiltonian $\hat{H}_0 = \omega_c \hat{a}^{\dagger}\hat{a} + E_0 \hat{\sigma}^{+}_{1}\hat{\sigma}^{-}_{1} + E_0 \hat{\sigma}^{+}_{2}\hat{\sigma}^{-}_{2} + + E_0 \hat{\sigma}^{+}_{3}\hat{\sigma}^{-}_{3} $, one can expand the wavefunction (\ref{psi3q}) and the Hamiltonian as shown in Eqs. (\ref{psi_exp}) and (\ref{H_exp}) respectively. Solving the set of differential equations (\ref{alpha_j}) obtained from the perturbative expansion of the Schroedinger equation, one can find the time-dependent coefficients $\alpha \left( t \right)$. The details of the derivation are presented in Appendix \ref{app_3q}.
Substituting the expression of the coefficients $\alpha \left( t \right)$ into the perturbative expansion up to second order in terms of $\frac{g_0}{2}$ of the wavefunction (\ref{psi3q}) and considering $n=0,1$ photons in the cavity, we obtain the following approximate solution of the Schroedinger equation

\begin{eqnarray}
\label{psi_exp2q3}
\begin{split}
{\lvert \psi \left( t \right)  \rangle} = {\lvert ggg,0  \rangle}^{(0)} + \left\{  \frac{g_0}{2} \frac{1}{\omega + E_0} \left( e^{-i \left( \omega + E_0 \right) t} -1 \right) \left[ {\lvert gge,1  \rangle}^{(1)} + {\lvert geg,1  \rangle}^{(1)} + {\lvert egg,1  \rangle}^{(1)} \right] \right\} + \\ 
+ \left\{  3 \frac{g_0^2}{4} \frac{1}{\left( \omega + E_0 \right)^2} \left( i \left( \omega + E_0 \right) t +  e^{-i \left( \omega + E_0 \right) t} -1 \right) {\lvert ggg,0  \rangle}^{(2)} + \right. \\
 \left. + \frac{g_0^2}{4}\frac{1}{E_0 \left( \omega + E_0 \right)^2 \left( E_0 -  \omega  \right)} \left[ 2E_0 - 2E_0 e^{-i \left( \omega + E_0 \right) t} + \left( \omega + E_0 \right) \left( e^{-i \left(  2E_0 \right) t} -1 \right) \right] {\lvert eee,0  \rangle}^{(2)}      \right\}.
\end{split}
\end{eqnarray}

\begin{figure}[t]
\subfloat[]{\label{fig:negativity_3q}}{\includegraphics[width = 2in, height = 1.5 in]{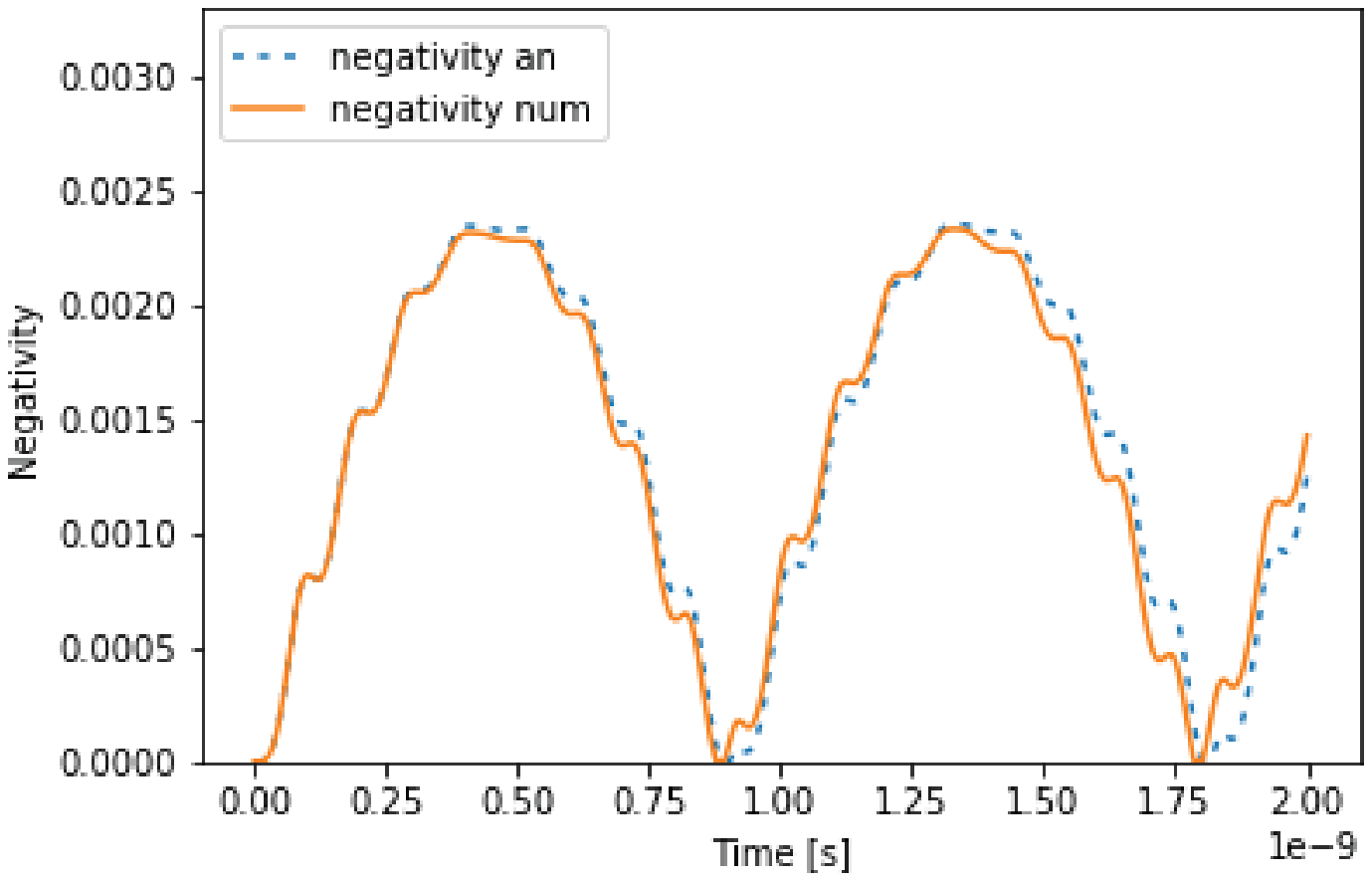}}
\subfloat[]{\label{fig:three_pi_3q}}{\includegraphics[width = 2 in, height = 1.5 in]{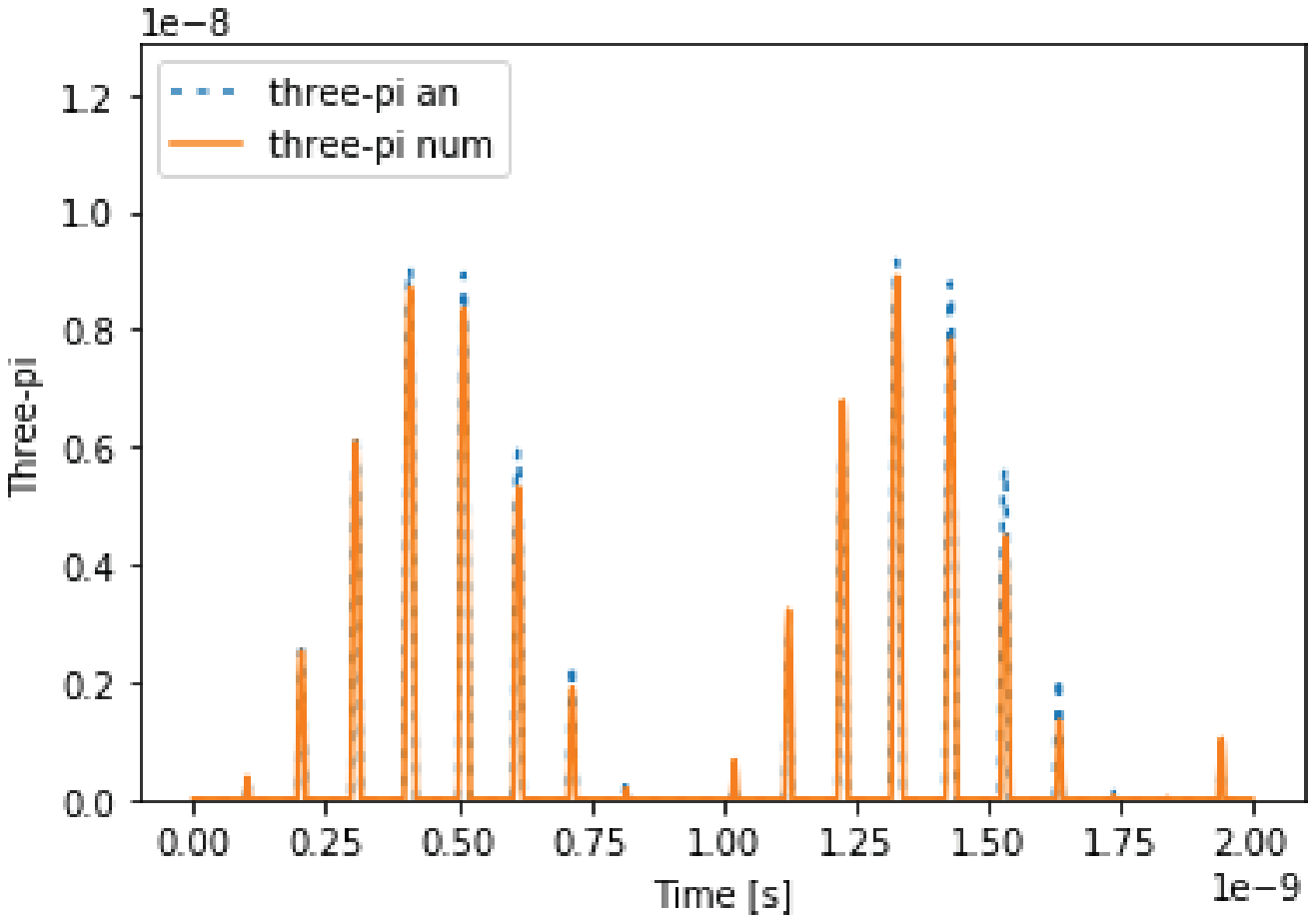}} 
\caption{Comparison of the analytical and numerical calculations for the time evolution of different measures of entanglement. (a) Negativity, and (b) three-$\pi$.}
\label{num_an_3q}
\end{figure}

\noindent
Using this analytical solution of Eq. (\ref{schroedinger}), one can calculate the measures of entanglement introduced in Sec. \ref{ment}. In particular, we substitute the wavefunction (\ref{psi_exp2q3}) into Eqs. (\ref{neg_abs}) and (\ref{3pi}) to calculate the negativity and the three-$\pi$, respectively.
At the same time,  Eq. (\ref{schroedinger}) is solved numerically with the same Hamiltonian and initial wavefunction. The values of the parameters used for the qubit and cavity frequencies and the qubit/cavity coupling strength are the same as the ones used in the corresponding subsection in the previous Section. The comparison between the time evolution of the different entanglement measures calculated using the perturbative analytical and numerical approaches are presented in Fig. \ref{num_an_3q}. As for the case of two qubits, there is excellent agreement between the perturbative and numerical calculations for both the entanglement measures used in Figs. \ref{fig:negativity_3q} and \ref{fig:three_pi_3q} and the two curves overlap almost perfectly.

\subsection{Dynamical Lamb effect with dissipation: numerical calculations}

Let us now consider the case where the qubit/cavity coupling is periodically switched-on/off nonadiabatically when dissipative effects are present. The instantaneous switching of the coupling leads to the excitation of the qubits and the creation of photons due to the dynamical Lamb effect. To study the time evolution of the quantum entanglement between the qubits under the driving of the coupling, we numerically solve Lindblad's master equation (\ref{lindblad}) for the system of three qubits coupled to a cavity that can be written as

\begin{equation}
\label{lindblad_3q}
\begin{split}
\frac{d \rho_{s} \left( t \right)  }{dt} = -i \left[ \hat{H}\left( t \right) , \rho_{s} \left( t \right) \right] 
+ \frac{k}{2} \left\{ 2 \hat{a}     \rho_{s} \left( t \right) \hat{a}^{\dagger} -  \rho_{s} \left( t \right) \hat{a}^{\dagger} \hat{a}  -  \hat{a}^{\dagger} \hat{a} \rho_{s} \left( t \right) \right\} +  \\
+ \sum_{i=1}^{3} \frac{\gamma_j}{2} \left\{ 2 \hat{\sigma}^{-}_{j}      \rho_{s} \left( t \right) \hat{\sigma}^{+}_{j} -  \rho_{s} \left( t \right) \hat{\sigma}^{+}_{j} \hat{\sigma}^{-}_{j} -  \hat{\sigma}^{+}_{j} \hat{\sigma}^{-}_{j} \rho_{s} \left( t \right) \right\} 
+ {\gamma_{\phi_j}} \left\{ \hat{\sigma}^{(3)}_{j}     \rho_{s} \left( t \right) \hat{\sigma}^{(3)}_{j} -  \rho_{s} \left( t \right)  \right\},
\end{split} 
\end{equation}

\noindent
where $\hat{H}\left( t \right)$ is the Hamiltonian (\ref{H3q}). In the numerical calculations, we use realistic values of the parameters of the system taken from the experiment done in Ref. \cite{lu} and specified in the previous Section. 

\begin{figure}[h]
\subfloat[]{\label{fig:negativity_dle_3q}}{\includegraphics[width = 2in, height = 1.5 in]{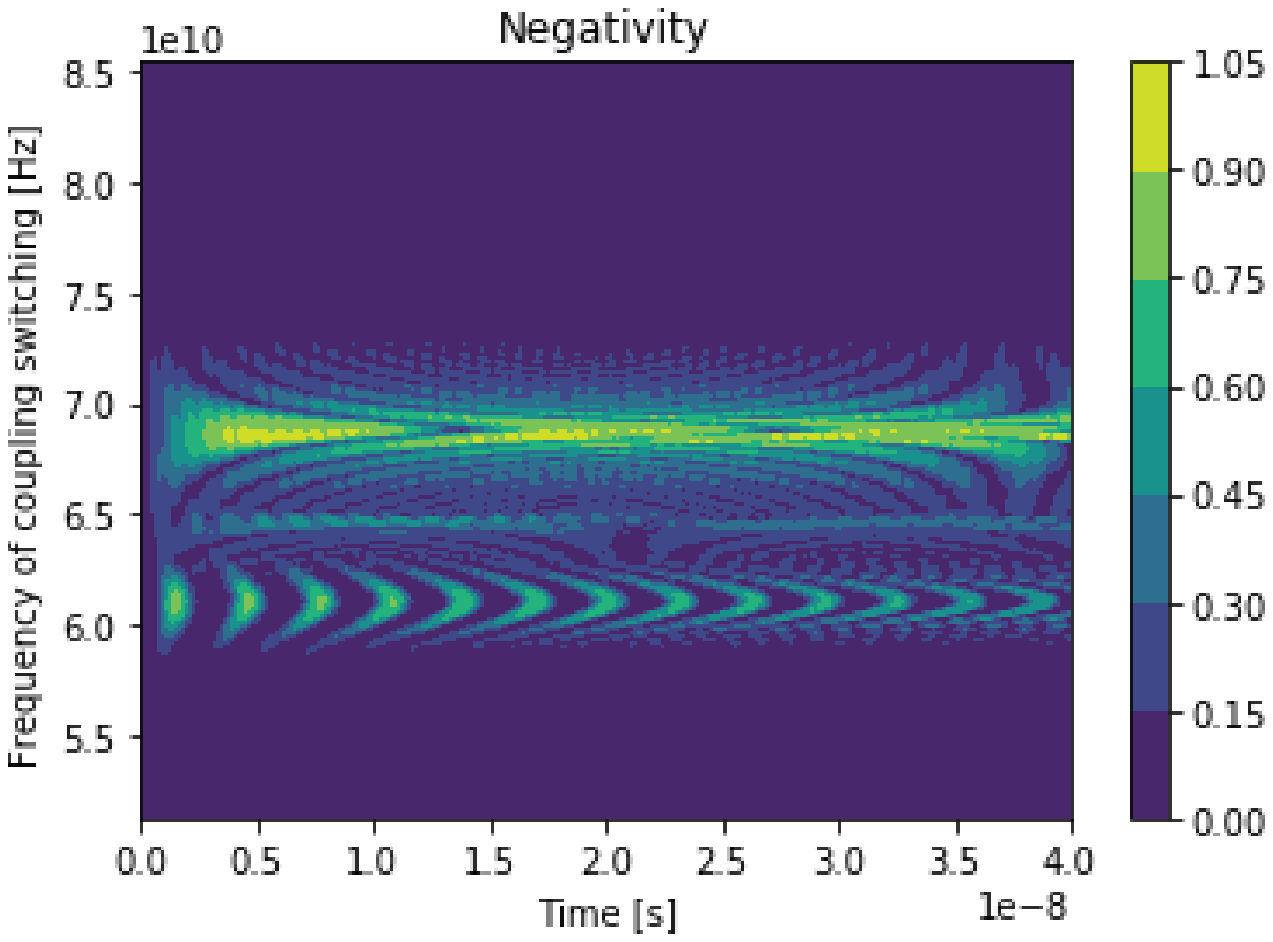}}
\subfloat[]{\label{fig:threepi_3q_dle}}{\includegraphics[width = 2 in, height = 1.5 in]{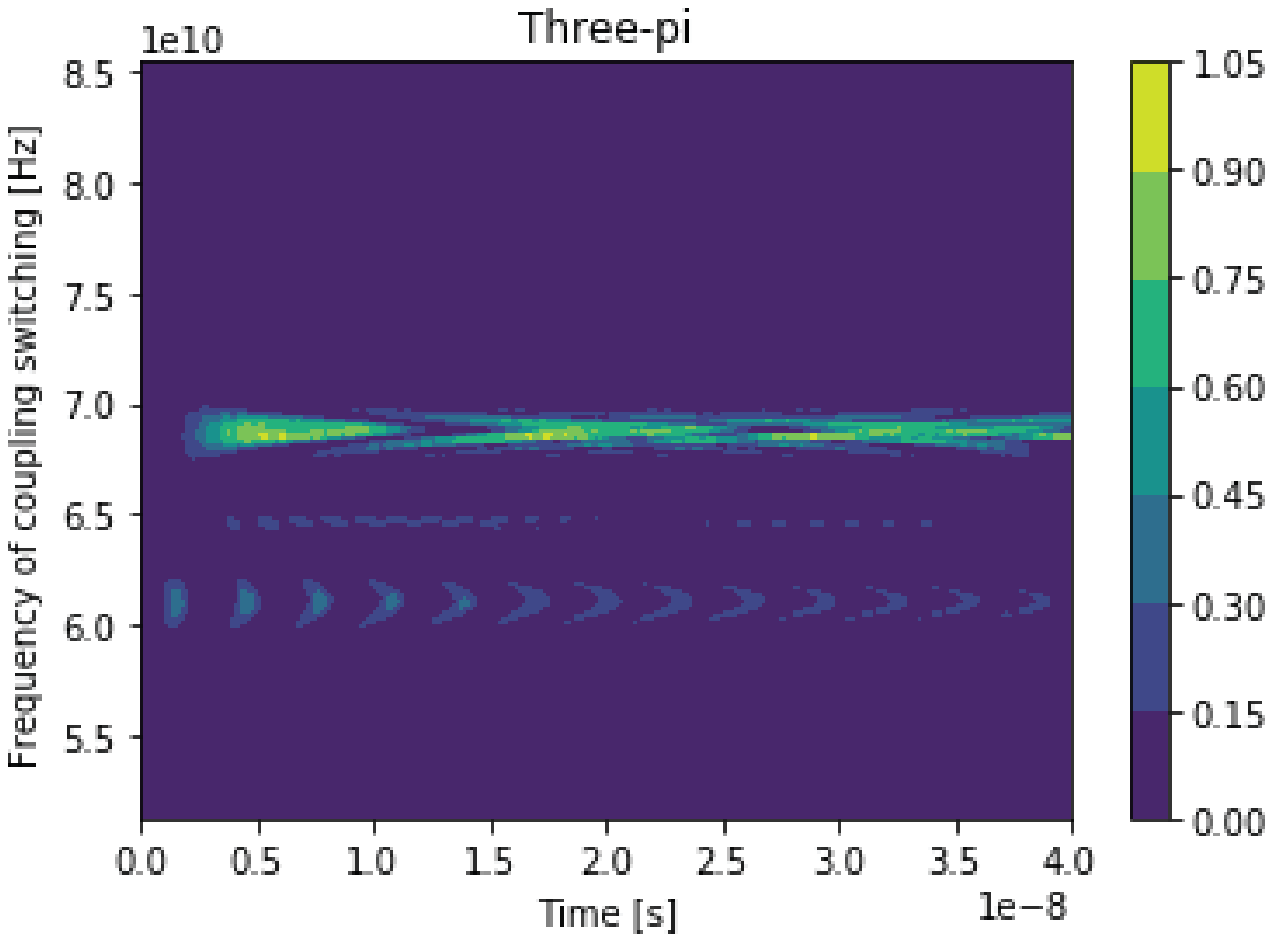}} 
\caption{Time evolution of different measures of entanglement for a range of frequencies of switching of the coupling. (a) Negativity, and (b) three-$\pi$.}
\label{span_f_3q}
\end{figure}

\begin{figure}[t]
\subfloat[]{\label{fig:negativity1_dle}}{\includegraphics[width = 2in, height = 1.5 in]{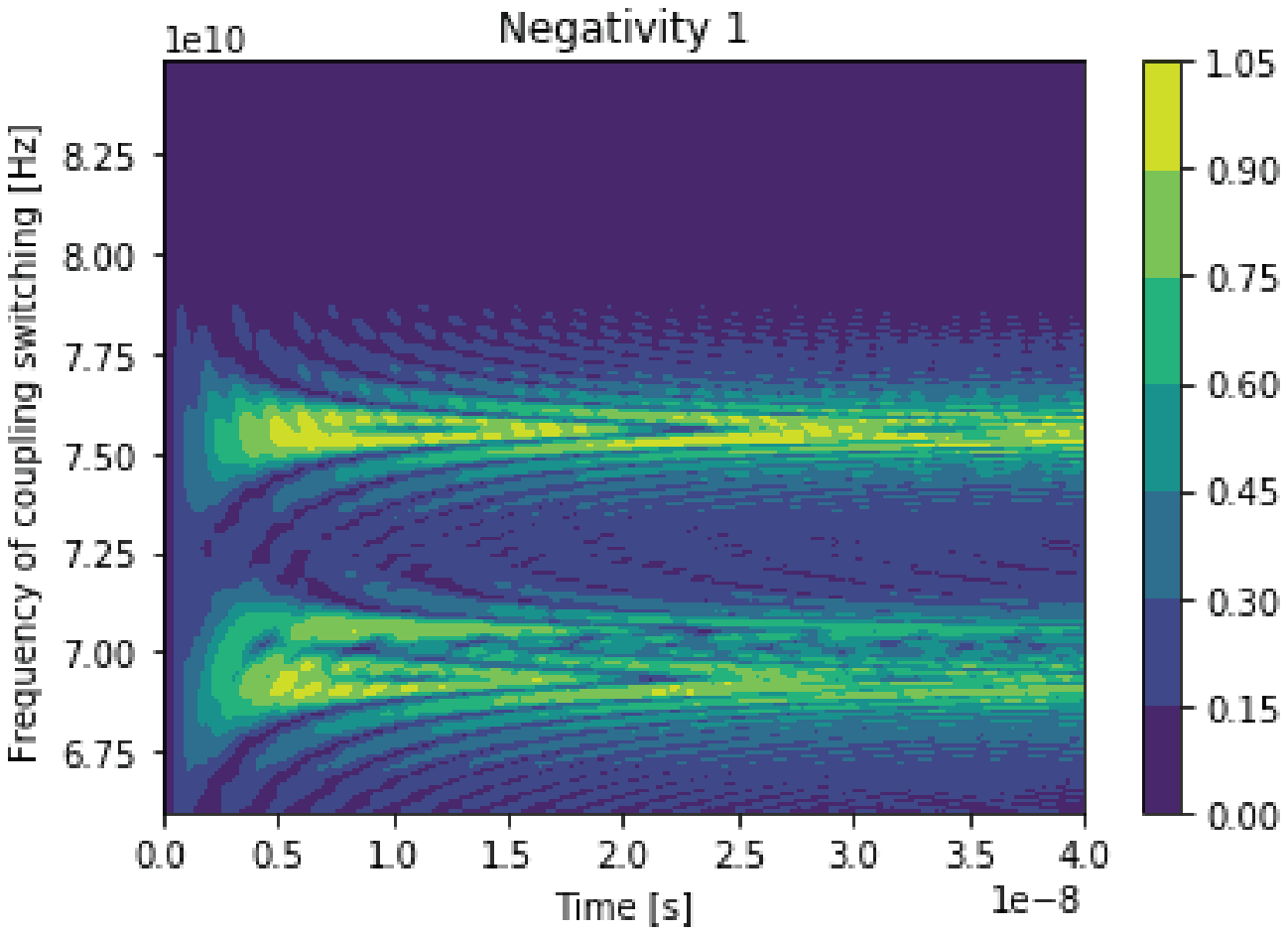}}
\subfloat[]{\label{fig:negativity2_dle}}{\includegraphics[width = 2in, height = 1.5 in]{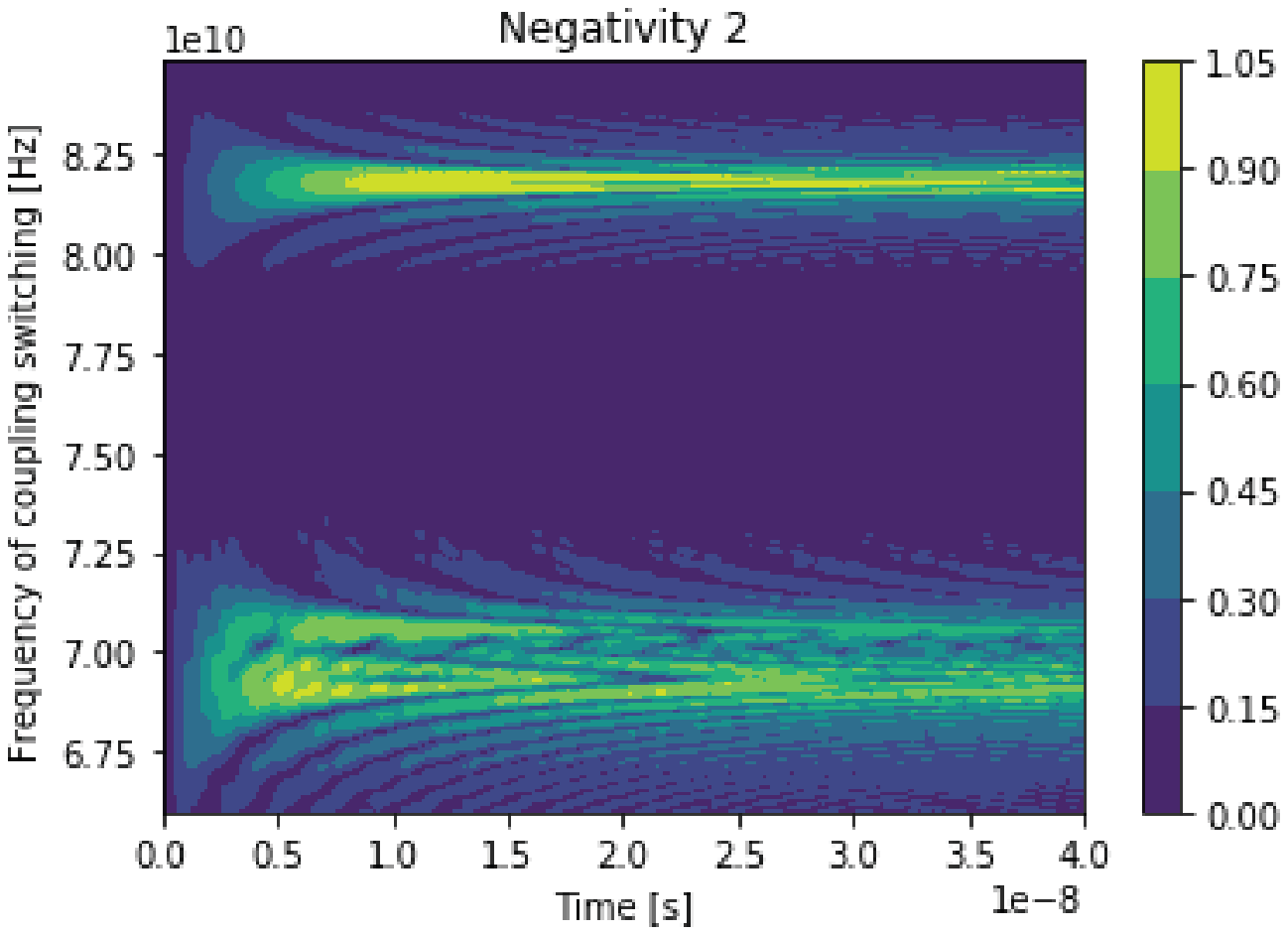}}
\subfloat[]{\label{fig:negativity3_dle}}{\includegraphics[width = 2in, height = 1.5 in]{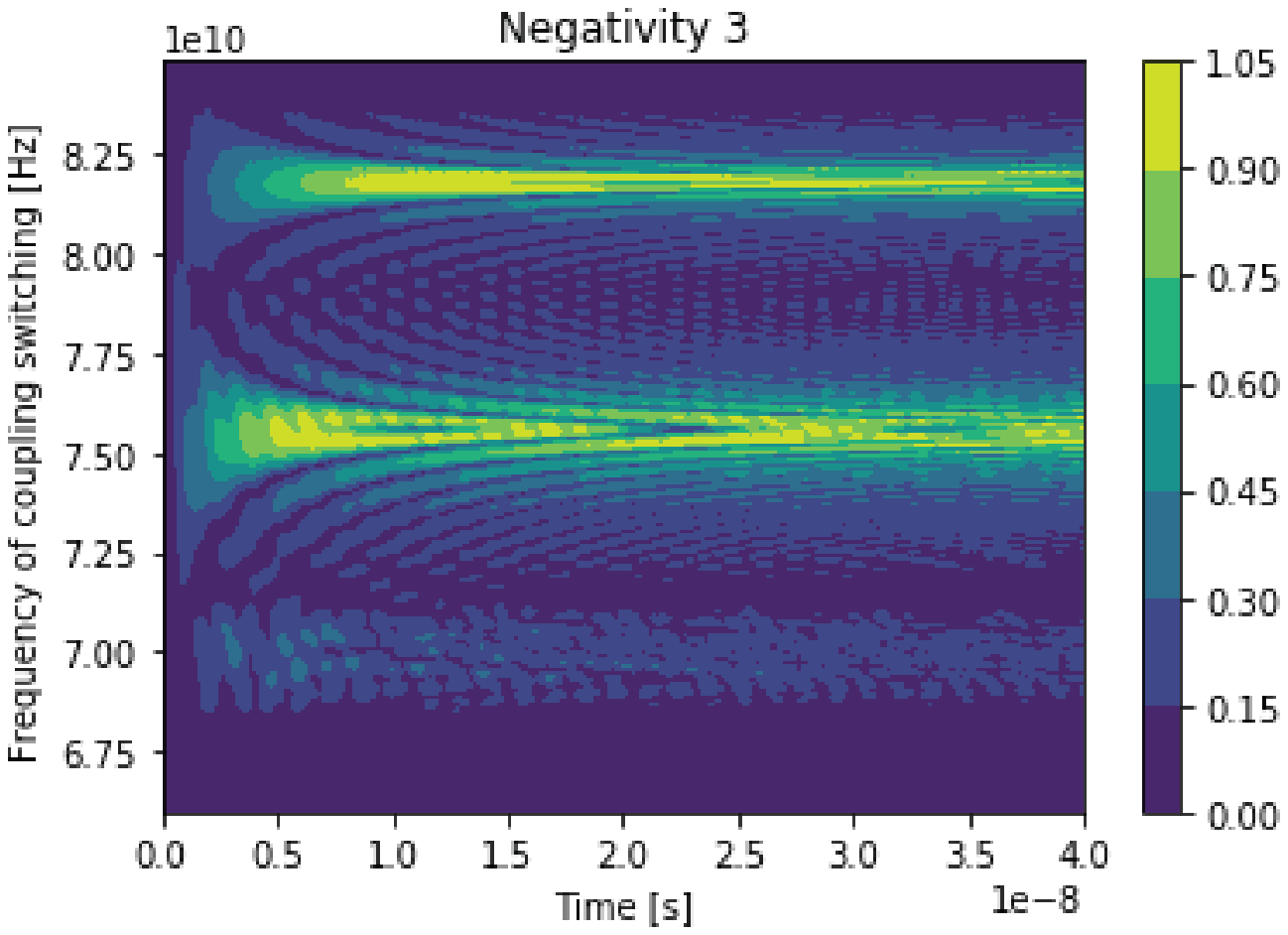}}
\subfloat[]{\label{fig:threepi_3q_diff_dle}}{\includegraphics[width = 2 in, height = 1.5 in]{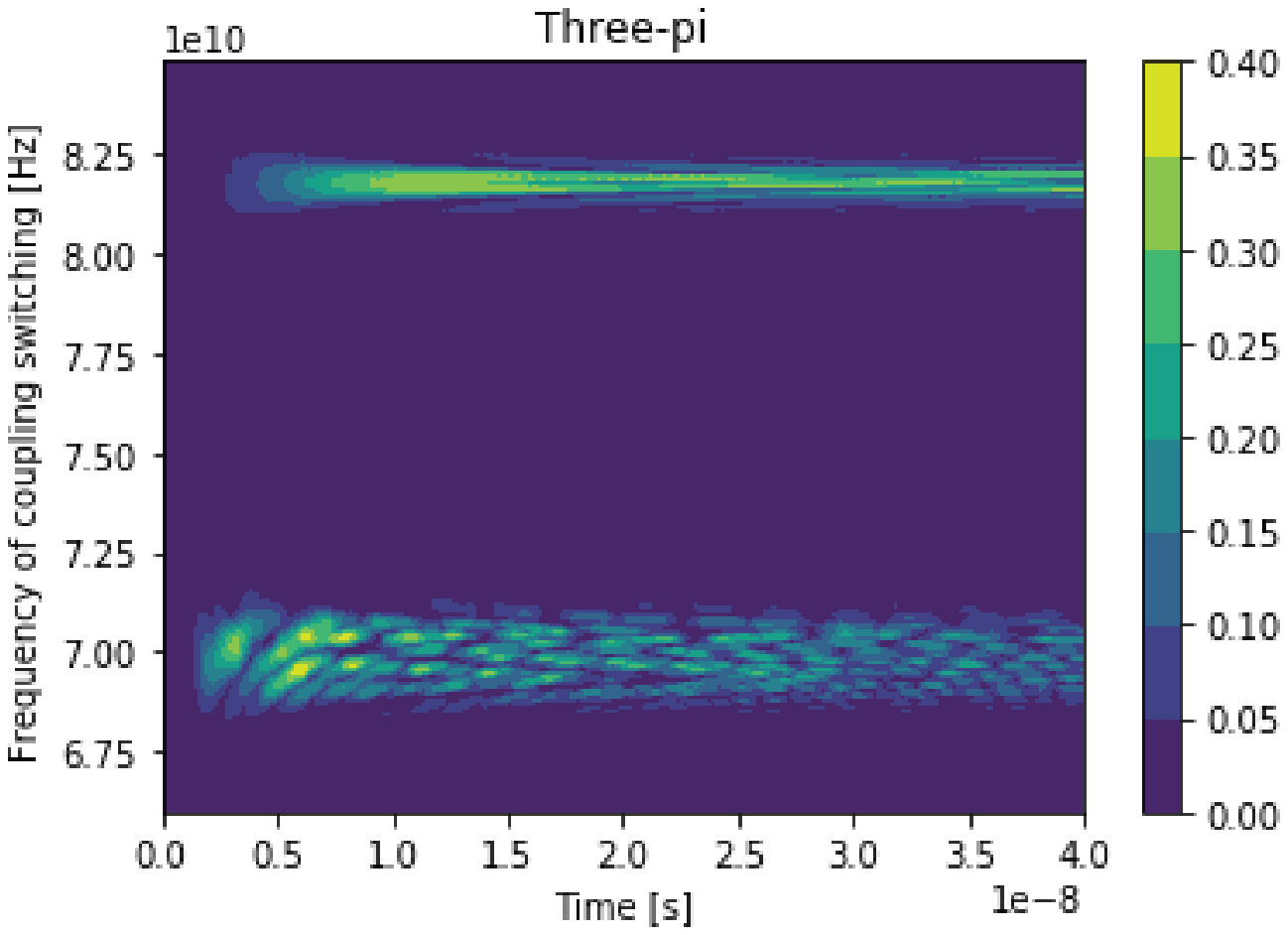}} 
\caption{Time evolution over a range of frequencies of switching of the coupling for (a) negativity of the first qubit, (b) negativity of the second qubit, (c) negativity of the third qubit, and (d) three-$\pi$.}
\label{ent2qub}
\end{figure}

The results of the calculations are shown in Figs. \ref{span_f_3q} - \ref{ent2qub}. In Figs. \ref{fig:negativity_dle_3q}, \ref{fig:threepi_3q_dle} the frequency of switching of the qubit/cavity coupling $\varpi_s$ is tuned over a certain range $\varpi_s \in \left[ E_0 , 4E_0 \right]$ to find the best value of this parameter which maximizes the entanglement between the qubits. 
We also consider the case where the qubits' transition frequencies are all different from each other. In particular, we choose $E_0^{(1)} =  2 \pi \times 5$ GHz, $E_0^{(2)} =  2 \pi \times 6$ GHz, $E_0^{(3)} =  2 \pi \times 7$ GHz. All other parameters are left unchanged.
By tuning the frequency $\varpi_s$ over the range $\varpi_s \in \left[ \frac{7}{4}E_0^{(2)} , \frac{9}{4}E_0^{(2)} \right]$, centered around twice the transition frequency of the second qubit, we find interesting features of the entanglement between the qubits that are discussed in Sec. \ref{results}. 

\section{Results and discussion}
\label{results}
In this Section we present and discuss the results of our calculations. The case of two and three qubits with time-independent coupling to the cavity is treated analytically and numerically in the absence of dissipation. 
It is clear from Figs. \ref{fig:concurrence} and \ref{fig:negativity} that the value of the concurrence coincide with the one of the negativity. Our calculations support the theoretical result obtained in Refs. \cite{verstraete, vidal2}. Because of the excellent agreement between approximate analytical calculations and numerical ones in Figs. \ref{num_an_2q} and \ref{num_an_3q}, we believe that the analytical method adopted can be successfully used whenever it's applicable. In the case considered, the approximation is valid for low frequency of switching of the coupling $\varpi_s \ll \omega_c + E_0$. However, when comparing the results obtained in the framework of the perturbative approach with the numerical calculations for the time-dependent coupling, it is clear that the approximation completely misses the picture that arises when the high frequency features are considered. 


For the two qubits case, it is evident from Figs. \ref{fig:concurrence_dle} and \ref{fig:negativity_dle} that there are two particular frequencies of switching of the coupling that maximizes the entanglement between the qubits. The first set of sharp bright peaks appears at the sum frequency of the transition frequencies of the qubit $\varpi_s = E_0^{(1)} + E_0^{(2)} = 2 E_0 $ in both measures of entanglement. One can also see another set of peaks around the sum of the qubit and cavity frequencies, $\varpi_s = \omega_c + E_0$. We believe that for this values of the frequency of switching the entanglement between the qubits is caused by the absorption of photons created through the DLE. This would explain the asymmetric nature of the fringes which appear only around the qubit/cavity sum frequency. It is interesting to note that the different measures of entanglement show different level of details: the concurrence in Fig. \ref{fig:concurrence_dle} is a sharper measure of the entanglement than the negativity, as it is markedly different from zero only at the values of the switching frequency that approach specific frequencies characteristic of the system, as the sum frequency of the cavity photons and the transition frequency of the qubit $\omega_c + E_0$ or the sum of the transition frequencies of two qubits $2E_0$. On the other hand, the negativity in Fig. \ref{fig:negativity_dle}, shows several peaks and fringes along with the same features of the concurrence. The fringed pattern here is much more visible and a more intricate structure appears also at frequencies different from the resonances of the system but it seems to fade with time.  The results in Fig. \ref{fig:negativity_dle} show that it is possible to realize an entangling gate between two qubits by turning off the nonadiabatic modulation of the coupling after a short time $t_0 \approx 7$ns. Figs. \ref{fig:concurrence_dle_phot}, \ref{fig:negativity_dle_phot} depict the results obtained for the time evolution of the quantum entanglement when the frequency of the cavity photons is tuned. The entanglement is degraded when the frequency of the cavity photons approaches the transition frequency of the qubits. When the qubit and the cavity are in resonance, the interaction between the qubits and the cavity photons destroys the entanglement between the qubits. In Figs. \ref{fig:concurrence_dle_diss} and \ref{fig:negativity_dle_diss}, the outcomes of the numerical calculations are presented. They show that for a cavity with high losses $\kappa / \omega_c \approx 0.1 $ one can generate steady state entanglement between the qubits, although the measures of entanglement do not reach their maximum. For lower cavity losses, the measures of entanglement between the qubits show features periodic in time, alternating between their maximal value and zero. Thus, one can engineer its system to achieve the desired characteristics. High cavity losses allow for steady state entanglement, while low cavity losses allow for fast, effective entangling gates. One can notice that dissipation and dephasing have a mild effect on the entanglement in the system if the qubit/cavity coupling is under modulation. The entanglement between the qubit due to the DLE is slowly damped due to the relaxation/dephasing of the qubit, while the entanglement between the qubit due to the Casimir photons is quickly damped. Thus, to preserve entanglement for longer times, one must consider a qubit with low decay/dephasing rate and a cavity with higher losses. The latter helps improve the lifetime of the entanglement between the qubits by decreasing qubit/photon interaction. 
Our findings support the results obtained in \cite{remizov}, where a more detailed analysis of the effect of photons in the cavity is carried out.

For the three qubits case, it is clear that there are two particular frequencies of switching of the coupling that maximize the entanglement between the qubits. These are indicated by the sharp bright peaks that appear in both measures of entanglement at the sum frequency of the transition frequencies of two qubits $\varpi_s = E_0^{(1)} + E_0^{(2)} = 2 E_0 $ and at the sum of the qubit and cavity frequencies, $\varpi_s = \omega_c + E_0$. 
The three-$\pi$ in Fig. \ref{fig:threepi_3q_dle} displays the entanglement between all three qubits, it is different from zero only at the values of the switching frequency that approach the sum frequency of the cavity photons and the transition frequency of the qubit $\omega_c + E_0$ or the sum of the transition frequencies of two qubits $2E_0$. Furthermore, since the three-$\pi$ measures the simultaneous entanglement of all the qubits, its value being close to one is an indication that GHZ states are produced when the system is driven at the qubits' sum frequency. The negativity in Fig. \ref{fig:negativity_dle_3q}, shows other peaks and fringes along with the features seen in the three-$\pi$. High value of the negativity appears around the qubits sum frequency and the qubit/cavity sum frequency. The fringed pattern here is much more visible, a more intricate structure appears also at frequencies different from the resonances of the system but it seems to quickly disappear. When the three qubits have the same transition frequency, as seen by the maximal value of the three-$\pi$ in Fig. \ref{fig:threepi_3q_dle}, it is possible to produce a maximally entangled GHZ state. From the GHZ theorem \cite{ghz, ghz2}, follows that a system of three entangled bodies can be used as a test for the validity of quantum mechanics. Thus, the proposed setup can be used to generate an entangled three qubit state and carry out such test. GHZ states can also be used as a way to implement the simplest quantum error correcting codes \cite{reed}, therefore providing a useful tool towards fault-tolerant quantum computation.
The results in Fig. \ref{fig:negativity3_dle} show the possibility of realizing two qubit gates if the transition frequencies of the qubits are different from each other. 
This is shown in Figs. \ref{fig:negativity1_dle}, \ref{fig:negativity2_dle}, \ref{fig:negativity3_dle}, where the negativity reaches its maximum value when the qubit/cavity coupling is driven at a frequency equal to the sum frequency of two qubits of the system. From Fig. \ref{fig:threepi_3q_diff_dle}, one can see that the three-$\pi$ never reaches its maximum value, indicating that the simultaneous entanglement of all three qubits is not the main channel of entanglement between them. 
Therefore, the dynamical Lamb effect can be used to selectively entangle two qubits connected through a shared bath by driving their coupling at the sum frequency of their transition frequencies. The high speed and degree of entanglement that can be achieved in this way, suggests that this could be a promising entangling gate, a fundamental building block to realize two qubits gate like the C-NOT gate. 

\section{Conclusions}
\label{conclusions}

We study the time evolution of the quantum entanglement generated by the dynamical Lamb effect between $N$ qubits coupled to a common resonator where dissipation is present. 
Following Refs. \cite{shapiro, zhukov, remizov}, we propose a physical realization of DLE driven quantum entanglement of $N$ superconducting qubits, whose coupling to a common resonator can be modulated through the use of auxillary SQUIDs. The use of SQUIDs to turn on/off the coupling allows to enter the nonadiabatic regime, where new quantum phenomena as the DLE and the DCE start to play an important role. However, all the physical realizations of superconducting systems with tunable coupling implemented up to now \cite{lu,mckay,roth} cause a shift of the qubit and cavity frequencies. Only recently, this issue was overcome by designing the qubit as a different arrangement of Josephson junctions \cite{he}.




We give a quantitative analysis of the $N=2$ and $3$ qubits cases under the assumptions of absence of dissipation and single switching of the coupling, which is then maintained constant over time. The time evolution of different measures of entanglement is calculated using the perturbative approach, that allows to find analytical solutions of Eq. (\ref{schroedinger}) and compare them with the purely numerical calculations. In the two qubits case, we use the concurrence, the mutual information and the negativity to measure the quantum entanglement in the system. In the three qubits case, we adopt the concurrence and the three-$\pi$.
Excellent agreement between the perturbative analytical calculations and the numerical ones is found in all cases at low frequencies, although the perturbative analytical approximation fails to correctly describe the effects caused by the high-frequency periodic switching of the coupling.


For this reason, we then consider a dissipative system of $N=2$ and $3$ qubits coupled to a common resonator, where the qubit/cavity coupling is suddenly switched on/off periodically. We investigate the dependence of several measures of quantum entanglement between the qubits on the parameters of the system to find the values which maximize the quantum entanglement between the qubits. For the case of two qubits, we use the concurrence and the negativity as measures of entanglement of the mixed states of the system, while we use the negativity and the three-$\pi$ for the three qubits case. Our numerical calculations indicate that the entanglement between the qubits is maximum when the following conditions are met: i. the frequency of the switching of the coupling $\varpi_s$ equals the sum frequency of the frequencies of the two qubits which are entangled; ii. the frequency of the cavity photons $\omega_c$ and the transition frequency of the qubits $E_0$ are not in resonance with each other; iii. the dissipation rate of the cavity photons is low.

We adopt different measures to quantify the quantum entanglement between the qubits in the various cases because each of them captures different level of details. In particular, the concurrence is able to distinctly detect the maximum of the entanglement. While the negativity shows in details where the entanglement can be nonzero, although not maximum.

\appendix

\section{Two qubits perturbative analytical calculations}
\label{app_2q}

For the case of two qubits, we have the wavefunction (\ref{psi2q}). At the zero-th order in terms of $\frac{g_0}{2}$, for $n=0,1$ photons in the cavity, the set of differential equations (\ref{alpha_0}) becomes

\begin{eqnarray}
\label{coeff2q00_app}
i \frac{d {\alpha}^{(0)}_{gg,0} }{dt} = 0 ,\nonumber \\ 
i \frac{d {\alpha}^{(0)}_{ge,0} }{dt} = E_0 {\alpha}^{(0)}_{ge,0}  ,\nonumber \\
i \frac{d {\alpha}^{(0)}_{eg,0} }{dt} = E_0 {\alpha}^{(0)}_{eg,0} ,\nonumber \\
i \frac{d {\alpha}^{(0)}_{ee,0} }{dt} = 2E_0 {\alpha}^{(0)}_{ee,0}  ,
\end{eqnarray}

\begin{eqnarray}
\label{coeff2q01_app}
i \frac{d {\alpha}^{(0)}_{gg,1} }{dt} = \omega {\alpha}^{(0)}_{gg,1} , \nonumber \\ 
i \frac{d {\alpha}^{(0)}_{ge,1} }{dt} = \left( \omega + E_0 \right) {\alpha}^{(0)}_{ge,1} , \nonumber \\
i \frac{d {\alpha}^{(0)}_{eg,1} }{dt} = \left( \omega + E_0 \right) {\alpha}^{(0)}_{eg,1} , \nonumber \\
i \frac{d {\alpha}^{(0)}_{ee,1} }{dt} = \left( \omega + 2E_0 \right) {\alpha}^{(0)}_{ee,1}.
\end{eqnarray}

\noindent
For the given initial condition, one finds that the only non-zero coefficient at zero-th order is ${\alpha}^{(0)}_{gg,0} = 1$.

At first order in terms of $\frac{g_0}{2}$ one finds 
 
\begin{eqnarray}
\label{coeff2q10_app}
i \frac{d {\alpha}^{(1)}_{gg,0} }{dt} =   {\alpha}^{(0)}_{ge,1} + {\alpha}^{(0)}_{eg,1}   ,\nonumber \\ 
i \frac{d {\alpha}^{(1)}_{ge,0} }{dt} = E_0 {\alpha}^{(1)}_{ge,0} +  {\alpha}^{(0)}_{gg,1} + {\alpha}^{(0)}_{ee,1}   ,\nonumber \\
i \frac{d {\alpha}^{(1)}_{eg,0} }{dt} = E_0 {\alpha}^{(1)}_{eg,0} + {\alpha}^{(0)}_{gg,1} + {\alpha}^{(0)}_{ee,1}  ,\nonumber \\
i \frac{d {\alpha}^{(1)}_{ee,0} }{dt} =  2E_0 {\alpha}^{(1)}_{ee,0} +   {\alpha}^{(0)}_{ge,1} + {\alpha}^{(0)}_{eg,1}   ,
\end{eqnarray}

\begin{eqnarray}
\label{coeff2q11_app}
i \frac{d {\alpha}^{(1)}_{gg,1} }{dt} = \omega{\alpha}^{(1)}_{gg,1} +   {\alpha}^{(0)}_{ge,0} + {\alpha}^{(0)}_{eg,0}   ,\nonumber \\ 
i \frac{d {\alpha}^{(1)}_{ge,1} }{dt} = \left( \omega + E_0 \right) {\alpha}^{(1)}_{ge,1} +  {\alpha}^{(0)}_{gg,0} + {\alpha}^{(0)}_{ee,0}    ,\nonumber \\
i \frac{d {\alpha}^{(1)}_{eg,1} }{dt} = \left( \omega + E_0 \right) {\alpha}^{(1)}_{eg,1} +    {\alpha}^{(0)}_{gg,0} + {\alpha}^{(0)}_{ee,0}   ,\nonumber \\
i \frac{d {\alpha}^{(1)}_{ee,1} }{dt} =  \left( \omega + 2E_0 \right) {\alpha}^{(1)}_{ee,1} +  {\alpha}^{(0)}_{ge,0} + {\alpha}^{(0)}_{eg,0}   .
\end{eqnarray}

\noindent
Substituting the value for the zero-th order coefficients $\alpha^{(0)}$, one can solve for the first order coefficients. The only non-zero coefficients at first order are

\begin{equation}
\label{coef2q1sol_app}
 {\alpha}^{(1)}_{ge,1} \left( t \right) = {\alpha}^{(1)}_{eg,1} \left( t \right) =   \frac{1}{\omega + E_0} \left( e^{-i \left( \omega + E_0 \right) t} -1 \right).
\end{equation}

At second order in terms of $\frac{g_0}{2}$, the set of differential equations (\ref{alpha_j}) reduces to

\begin{eqnarray}
\label{coeff2q20_app}
i \frac{d {\alpha}^{(2)}_{gg,0} }{dt} =   {\alpha}^{(1)}_{ge,1} + {\alpha}^{(1)}_{eg,1}  ,\nonumber \\ 
i \frac{d {\alpha}^{(2)}_{ge,0} }{dt} = E_0 {\alpha}^{(2)}_{ge,0} +     {\alpha}^{(1)}_{gg,1} + {\alpha}^{(1)}_{ee,1}     ,\nonumber \\
i \frac{d {\alpha}^{(2)}_{eg,0} }{dt} = E_0 {\alpha}^{(2)}_{eg,0} +     {\alpha}^{(1)}_{gg,1} + {\alpha}^{(1)}_{ee,1}    ,\nonumber \\
i \frac{d {\alpha}^{(2)}_{ee,0} }{dt} =  2E_0 {\alpha}^{(2)}_{ee,0} +    {\alpha}^{(1)}_{ge,1} + {\alpha}^{(1)}_{eg,1}    ,
\end{eqnarray}

\begin{eqnarray}
\label{coeff2q21_app}
i \frac{d {\alpha}^{(2)}_{gg,1} }{dt} = \omega{\alpha}^{(2)}_{gg,1} +     {\alpha}^{(1)}_{ge,0} + {\alpha}^{(1)}_{eg,0}    ,\nonumber \\ 
i \frac{d {\alpha}^{(2)}_{ge,1} }{dt} = \left( \omega + E_0 \right) {\alpha}^{(2)}_{ge,1} +     {\alpha}^{(1)}_{gg,0} + {\alpha}^{(1)}_{ee,0}     ,\nonumber \\
i \frac{d {\alpha}^{(2)}_{eg,1} }{dt} = \left( \omega + E_0 \right) {\alpha}^{(2)}_{eg,1} +     {\alpha}^{(1)}_{gg,0} + {\alpha}^{(1)}_{ee,0}    ,\nonumber \\
i \frac{d {\alpha}^{(2)}_{ee,1} }{dt} =  \left( \omega + 2E_0 \right) {\alpha}^{(2)}_{ee,1} +     {\alpha}^{(1)}_{ge,0} + {\alpha}^{(1)}_{eg,0}    .
\end{eqnarray}

\noindent
Substituting the value for the first order coefficients $\alpha^{(1)}$, one can find the second order coefficients. The only non-zero coefficients are the following

\begin{eqnarray}
\label{coef2q2sol_app}
 {\alpha}^{(2)}_{gg,0} \left( t \right)  &=& \frac{2}{\left( \omega + E_0 \right)^2} \left( i \left( \omega + E_0 \right) t +  e^{-i \left( \omega + E_0 \right) t} -1 \right), \nonumber \\
 {\alpha}^{(2)}_{ee,0} \left( t \right) &=&  \frac{1}{E_0 \left( \omega + E_0 \right)^2 \left( E_0 -  \omega  \right)} \left[ 2E_0 - 2E_0 e^{-i \left( \omega + E_0 \right) t} + \left( \omega + E_0 \right) \left( e^{-i \left(  2E_0 \right) t} -1 \right) \right] .
\end{eqnarray}

\noindent
Therefore, truncating the perturbative expansion of the wavefunction (\ref{psi2q}) at second order and substituting the value for $\alpha (t)$ we obtain the following approximate solution of the Schroedinger equation

\begin{eqnarray}
\label{psi_exp2q2_app}
\begin{split}
{\lvert \psi \left( t \right)  \rangle} =  {\lvert gg,0  \rangle}^{(0)} + \left\{  \frac{g_0}{2} \frac{1}{\omega + E_0} \left( e^{-i \left( \omega + E_0 \right) t} -1 \right) \left[ {\lvert ge,1  \rangle}^{(1)} + {\lvert eg,1  \rangle}^{(1)} \right] \right\} + \\ 
+ \left\{ \frac{g_0^2}{2} \frac{1}{\left( \omega + E_0 \right)^2} \left( i \left( \omega + E_0 \right) t +  e^{-i \left( \omega + E_0 \right) t} -1 \right) {\lvert gg,0  \rangle}^{(2)} + \right. \\
 \left. + \frac{g_0^2}{4}\frac{1}{E_0 \left( \omega + E_0 \right)^2 \left( E_0 -  \omega  \right)} \left[ 2E_0 - 2E_0 e^{-i \left( \omega + E_0 \right) t} + \left( \omega + E_0 \right) \left( e^{-i \left(  2E_0 \right) t} -1 \right) \right] {\lvert ee,0  \rangle}^{(2)}      \right\}.
\end{split}
\end{eqnarray}

\section{Three qubits perturbative analytical calculations}
\label{app_3q}

The same steps done in the two qubits case also apply for the case of three qubits. One starts with the wavefunction  (\ref{psi3q}) and solves the Schroedinger equation perturbatively as highlighted in Sec. \ref{3q}.

At the zero-th order in terms of $\frac{g_0}{2}$, for the case of $n=0,1$ photons in the cavity, this translates into the following differential equations for the time-dependent coefficients $\alpha$

\begin{eqnarray}
\label{coeff3q00_app}
i \frac{d {\alpha}^{(0)}_{ggg,0} }{dt} = 0 ,\nonumber \\ 
i \frac{d {\alpha}^{(0)}_{gge,0} }{dt} = E_0 {\alpha}^{(0)}_{gge,0}  ,\nonumber \\
i \frac{d {\alpha}^{(0)}_{geg,0} }{dt} = E_0 {\alpha}^{(0)}_{geg,0}  ,\nonumber \\
i \frac{d {\alpha}^{(0)}_{egg,0} }{dt} = E_0 {\alpha}^{(0)}_{egg,0}  ,\nonumber \\
i \frac{d {\alpha}^{(0)}_{eeg,0} }{dt} = 2E_0 {\alpha}^{(0)}_{eeg,0} ,\nonumber \\
i \frac{d {\alpha}^{(0)}_{ege,0} }{dt} = 2E_0 {\alpha}^{(0)}_{ege,0} ,\nonumber \\
i \frac{d {\alpha}^{(0)}_{gee,0} }{dt} = 2E_0 {\alpha}^{(0)}_{gee,0} ,\nonumber \\
i \frac{d {\alpha}^{(0)}_{eee,0} }{dt} = 3E_0 {\alpha}^{(0)}_{eee,0}  ,
\end{eqnarray}

\begin{eqnarray}
\label{coeff2q01_app}
i \frac{d {\alpha}^{(0)}_{ggg,1} }{dt} = \omega {\alpha}^{(0)}_{ggg,1} , \nonumber \\ 
i \frac{d {\alpha}^{(0)}_{gge,1} }{dt} = \left( \omega + E_0 \right) {\alpha}^{(0)}_{gge,1} , \nonumber \\
i \frac{d {\alpha}^{(0)}_{geg,1} }{dt} = \left( \omega + E_0 \right) {\alpha}^{(0)}_{geg,1} , \nonumber \\
i \frac{d {\alpha}^{(0)}_{egg,1} }{dt} = \left( \omega + E_0 \right) {\alpha}^{(0)}_{egg,1} , \nonumber \\
i \frac{d {\alpha}^{(0)}_{eeg,1} }{dt} = \left( \omega + 2E_0 \right) {\alpha}^{(0)}_{eeg,1} , \nonumber \\
i \frac{d {\alpha}^{(0)}_{ege,1} }{dt} = \left( \omega + 2E_0 \right) {\alpha}^{(0)}_{ege,1} , \nonumber \\
i \frac{d {\alpha}^{(0)}_{gee,1} }{dt} = \left( \omega + 2E_0 \right) {\alpha}^{(0)}_{gee,1} , \nonumber \\
i \frac{d {\alpha}^{(0)}_{eee,1} }{dt} = \left( \omega + 3E_0 \right) {\alpha}^{(0)}_{eee,1}.
\end{eqnarray}

\noindent
For the given initial condition, one finds that the only non-zero coefficient is ${\alpha}^{(0)}_{ggg,0} = 1$.

At first order in terms of $\frac{g_0}{2}$ one obtains
 
\begin{eqnarray}
\label{coeff3q10_app}
i \frac{d {\alpha}^{(1)}_{ggg,0} }{dt} =  {\alpha}^{(0)}_{gge,1} + {\alpha}^{(0)}_{geg,1} + {\alpha}^{(0)}_{gge,1}  ,\nonumber \\ 
i \frac{d {\alpha}^{(1)}_{gge,0} }{dt} = E_0 {\alpha}^{(1)}_{gge,0} +  {\alpha}^{(0)}_{ggg,1} + {\alpha}^{(0)}_{gee,1} + {\alpha}^{(0)}_{ege,1}    ,\nonumber \\
i \frac{d {\alpha}^{(1)}_{geg,0} }{dt} = E_0 {\alpha}^{(1)}_{geg,0} + \ {\alpha}^{(0)}_{ggg,1} + {\alpha}^{(0)}_{gee,1} + {\alpha}^{(0)}_{eeg,1}    ,\nonumber \\
i \frac{d {\alpha}^{(1)}_{egg,0} }{dt} = E_0 {\alpha}^{(1)}_{egg,0} +  {\alpha}^{(0)}_{ggg,1} + {\alpha}^{(0)}_{ege,1} + {\alpha}^{(0)}_{egg,1}    ,\nonumber \\
i \frac{d {\alpha}^{(1)}_{eeg,0} }{dt} = 2E_0 {\alpha}^{(1)}_{eeg,0} +  {\alpha}^{(0)}_{geg,1} + {\alpha}^{(0)}_{egg,1} + {\alpha}^{(0)}_{eee,1}   ,\nonumber \\
i \frac{d {\alpha}^{(1)}_{ege,0} }{dt} = 2E_0 {\alpha}^{(1)}_{ege,0} +  {\alpha}^{(0)}_{gge,1} + {\alpha}^{(0)}_{egg,1} + {\alpha}^{(0)}_{eee,1}    ,\nonumber \\
i \frac{d {\alpha}^{(1)}_{gee,0} }{dt} = 2E_0 {\alpha}^{(1)}_{gee,0} +{\alpha}^{(0)}_{geg,1} + {\alpha}^{(0)}_{gge,1} + {\alpha}^{(0)}_{eee,1}   ,\nonumber \\
i \frac{d {\alpha}^{(1)}_{eee,0} }{dt} =  3E_0 {\alpha}^{(1)}_{eee,0} + {\alpha}^{(0)}_{eeg,1} + {\alpha}^{(0)}_{ege,1} + {\alpha}^{(0)}_{eeg,1}   ,
\end{eqnarray}

\begin{eqnarray}
\label{coeff3q11_app}
i \frac{d {\alpha}^{(1)}_{ggg,1} }{dt} = \omega {\alpha}^{(1)}_{ggg,1} +  {\alpha}^{(0)}_{gge,0} + {\alpha}^{(0)}_{geg,0} + {\alpha}^{(0)}_{gge,0}  ,\nonumber \\ 
i \frac{d {\alpha}^{(1)}_{gge,1} }{dt} = \left( \omega + E_0 \right) {\alpha}^{(1)}_{gge,1} +  {\alpha}^{(0)}_{ggg,0} + {\alpha}^{(0)}_{gee,0} + {\alpha}^{(0)}_{ege,0}    ,\nonumber \\
i \frac{d {\alpha}^{(1)}_{geg,1} }{dt} = \left( \omega + E_0 \right) {\alpha}^{(1)}_{geg,1} +  {\alpha}^{(0)}_{ggg,0} + {\alpha}^{(0)}_{gee,0} + {\alpha}^{(0)}_{eeg,0}    ,\nonumber \\
i \frac{d {\alpha}^{(1)}_{egg,1} }{dt} = \left( \omega + E_0 \right) {\alpha}^{(1)}_{egg,1} +  {\alpha}^{(0)}_{ggg,0} + {\alpha}^{(0)}_{ege,0} + {\alpha}^{(0)}_{egg,0}    ,\nonumber \\
i \frac{d {\alpha}^{(1)}_{eeg,1} }{dt} = \left( \omega + 2E_0 \right) {\alpha}^{(1)}_{eeg,1} + {\alpha}^{(0)}_{geg,0} + {\alpha}^{(0)}_{egg,0} + {\alpha}^{(0)}_{eee,0}  ,\nonumber \\
i \frac{d {\alpha}^{(1)}_{ege,1} }{dt} = \left( \omega + 2E_0 \right) {\alpha}^{(1)}_{ege,1} +  {\alpha}^{(0)}_{gge,0} + {\alpha}^{(0)}_{egg,0} + {\alpha}^{(0)}_{eee,0}    ,\nonumber \\
i \frac{d {\alpha}^{(1)}_{gee,1} }{dt} = \left( \omega + 2E_0 \right) {\alpha}^{(1)}_{gee,1} +  {\alpha}^{(0)}_{geg,0} + {\alpha}^{(0)}_{gge,0} + {\alpha}^{(0)}_{eee,0}   ,\nonumber \\
i \frac{d {\alpha}^{(1)}_{eee,1} }{dt} =  \left( \omega + 3E_0 \right) {\alpha}^{(1)}_{eee,1} + {\alpha}^{(0)}_{eeg,0} + {\alpha}^{(0)}_{ege,0} + {\alpha}^{(0)}_{eeg,0}  .
\end{eqnarray}

\noindent
The only non-zero coefficients at first order are

\begin{equation}
\label{coef3q1sol_app}
 {\alpha}^{(1)}_{gge,1} \left( t \right) = {\alpha}^{(1)}_{geg,1} \left( t \right) = {\alpha}^{(1)}_{egg,1} \left( t \right) = \frac{1}{\omega + E_0} \left( e^{-i \left( \omega + E_0 \right) t} -1 \right).
\end{equation}

At second order in terms of $\frac{g_0}{2}$, one gets a set of equations for the time-dependent coefficients

\begin{eqnarray}
\label{coeff3q20_app}
i \frac{d {\alpha}^{(2)}_{ggg,0} }{dt} =  {\alpha}^{(1)}_{gge,1} + {\alpha}^{(1)}_{geg,1} + {\alpha}^{(1)}_{gge,1}  ,\nonumber \\ 
i \frac{d {\alpha}^{(2)}_{gge,0} }{dt} = E_0 {\alpha}^{(2)}_{gge,0} +  {\alpha}^{(1)}_{ggg,1} + {\alpha}^{(1)}_{gee,1} + {\alpha}^{(1)}_{ege,1}    ,\nonumber \\
i \frac{d {\alpha}^{(2)}_{geg,0} }{dt} = E_0 {\alpha}^{(2)}_{geg,0} +  {\alpha}^{(1)}_{ggg,1} + {\alpha}^{(1)}_{gee,1} + {\alpha}^{(1)}_{eeg,1}   ,\nonumber \\
i \frac{d {\alpha}^{(2)}_{egg,0} }{dt} = E_0 {\alpha}^{(2)}_{egg,0} +   {\alpha}^{(1)}_{ggg,1} + {\alpha}^{(1)}_{ege,1} + {\alpha}^{(1)}_{egg,1}   ,\nonumber \\
i \frac{d {\alpha}^{(2)}_{eeg,0} }{dt} = 2E_0 {\alpha}^{(2)}_{eeg,0} +   {\alpha}^{(1)}_{geg,1} + {\alpha}^{(1)}_{egg,1} + {\alpha}^{(1)}_{eee,1}   ,\nonumber \\
i \frac{d {\alpha}^{(2)}_{ege,0} }{dt} = 2E_0 {\alpha}^{(2)}_{ege,0} +   {\alpha}^{(1)}_{gge,1} + {\alpha}^{(1)}_{egg,1} + {\alpha}^{(1)}_{eee,1}   ,\nonumber \\
i \frac{d {\alpha}^{(2)}_{gee,0} }{dt} = 2E_0 {\alpha}^{(2)}_{gee,0} +   {\alpha}^{(1)}_{geg,1} + {\alpha}^{(1)}_{gge,1} + {\alpha}^{(1)}_{eee,1}   ,\nonumber \\
i \frac{d {\alpha}^{(2)}_{eee,0} }{dt} =  3E_0 {\alpha}^{(2)}_{eee,0} +  {\alpha}^{(1)}_{eeg,1} + {\alpha}^{(1)}_{ege,1} + {\alpha}^{(1)}_{eeg,1}  ,
\end{eqnarray}

\begin{eqnarray}
\label{coeff3q21_app}
i \frac{d {\alpha}^{(2)}_{ggg,1} }{dt} = \omega {\alpha}^{(2)}_{ggg,1} +   {\alpha}^{(1)}_{gge,0} + {\alpha}^{(1)}_{geg,0} + {\alpha}^{(1)}_{gge,0}  ,\nonumber \\ 
i \frac{d {\alpha}^{(2)}_{gge,1} }{dt} = \left( \omega + E_0 \right) {\alpha}^{(2)}_{gge,1} +   {\alpha}^{(1)}_{ggg,0} + {\alpha}^{(1)}_{gee,0} + {\alpha}^{(1)}_{ege,0}   ,\nonumber \\
i \frac{d {\alpha}^{(2)}_{geg,1} }{dt} = \left( \omega + E_0 \right) {\alpha}^{(2)}_{geg,1} +   {\alpha}^{(1)}_{ggg,0} + {\alpha}^{(1)}_{gee,0} + {\alpha}^{(1)}_{eeg,0}   ,\nonumber \\
i \frac{d {\alpha}^{(2)}_{egg,1} }{dt} = \left( \omega + E_0 \right) {\alpha}^{(2)}_{egg,1} +   {\alpha}^{(1)}_{ggg,0} + {\alpha}^{(1)}_{ege,0} + {\alpha}^{(1)}_{egg,0}   ,\nonumber \\
i \frac{d {\alpha}^{(2)}_{eeg,1} }{dt} = \left( \omega + 2E_0 \right) {\alpha}^{(2)}_{eeg,1} +   {\alpha}^{(1)}_{geg,0} + {\alpha}^{(1)}_{egg,0} + {\alpha}^{(1)}_{eee,0}    ,\nonumber \\
i \frac{d {\alpha}^{(2)}_{ege,1} }{dt} = \left( \omega + 2E_0 \right) {\alpha}^{(2)}_{ege,1} +   {\alpha}^{(1)}_{gge,0} + {\alpha}^{(1)}_{egg,0} + {\alpha}^{(1)}_{eee,0}    ,\nonumber \\
i \frac{d {\alpha}^{(2)}_{gee,1} }{dt} = \left( \omega + 2E_0 \right) {\alpha}^{(2)}_{gee,1} +   {\alpha}^{(1)}_{geg,0} + {\alpha}^{(1)}_{gge,0} + {\alpha}^{(1)}_{eee,0}   ,\nonumber \\
i \frac{d {\alpha}^{(2)}_{eee,1} }{dt} =  \left( \omega + 3E_0 \right) {\alpha}^{(2)}_{eee,1} +  {\alpha}^{(1)}_{eeg,0} + {\alpha}^{(1)}_{ege,0} + {\alpha}^{(1)}_{eeg,0}  .
\end{eqnarray}

\noindent
Here the only non-zero coefficients are

\begin{eqnarray}
\label{coef3q2sol_app}
 {\alpha}^{(2)}_{ggg,0} \left( t \right)  &=&  \frac{3}{\left( \omega + E_0 \right)^2} \left( i \left( \omega + E_0 \right) t +  e^{-i \left( \omega + E_0 \right) t} -1 \right), \nonumber \\
 {\alpha}^{(2)}_{eee,0} \left( t \right) &=&  \frac{1}{E_0 \left( \omega + E_0 \right)^2 \left( E_0 -  \omega  \right)} \left[ 2E_0 - 2E_0 e^{-i \left( \omega + E_0 \right) t} + \left( \omega + E_0 \right) \left( e^{-i \left(  2E_0 \right) t} -1 \right) \right] .
\end{eqnarray}

\noindent
Thus, truncating the expansion at the second order, we obtain the following solution to the Schroedinger equation

\begin{eqnarray}
\label{psi_exp3q2_app}
\begin{split}
{\lvert \psi \left( t \right)  \rangle} = {\lvert ggg,0  \rangle}^{(0)} + \left\{  \frac{g_0}{2} \frac{1}{\omega + E_0} \left( e^{-i \left( \omega + E_0 \right) t} -1 \right) \left[ {\lvert gge,1  \rangle}^{(1)} + {\lvert geg,1  \rangle}^{(1)} + {\lvert egg,1  \rangle}^{(1)} \right] \right\} + \\ 
+ \left\{  3 \frac{g_0^2}{4} \frac{1}{\left( \omega + E_0 \right)^2} \left( i \left( \omega + E_0 \right) t +  e^{-i \left( \omega + E_0 \right) t} -1 \right) {\lvert ggg,0  \rangle}^{(2)} + \right. \\
 \left. + \frac{g_0^2}{4}\frac{1}{E_0 \left( \omega + E_0 \right)^2 \left( E_0 -  \omega  \right)} \left[ 2E_0 - 2E_0 e^{-i \left( \omega + E_0 \right) t} + \left( \omega + E_0 \right) \left( e^{-i \left(  2E_0 \right) t} -1 \right) \right] {\lvert eee,0  \rangle}^{(2)}      \right\}.
\end{split}
\end{eqnarray}

\end{document}